\documentclass[12pt]{article}
\begin{document}

\begin{center} {\LARGE \bf Heart Rate Variability: Measures and Models}
\end{center}

\begin{center}
Malvin~C.~Teich, Steven~B.~Lowen, Bradley~M.~Jost, and
Karin~Vibe-Rheymer \\
Department of Electrical and Computer Engineering,
Boston University \\
8 Saint Mary's Street, Boston, MA 02215 \\
and \\
Conor~Heneghan\\
Department of Electronic and Electrical
Engineering, \\
University College Dublin, Belfield, Dublin 4,
Ireland
\end{center}

\pagebreak
\tableofcontents
\pagebreak

\section{Introduction}

The human heart generates the quintessential biological
signal: the heartbeat.  A recording of the cardiac-induced skin
potentials at the body's surface, an electrocardiogram
(ECG), reveals information about atrial and ventricular electrical
activity. Abnormalities in the temporal durations of the segments
between deflections, or of the intervals between waves in the ECG, as
well as their
relative heights, serve to expose and distinguish cardiac
dysfunction. Because the electrical activity of the human heart is
influenced by many physiological mechanisms, electrocardiography has
become an invaluable tool for the diagnosis of a variety of
pathologies that affect the cardiovascular system \cite{levick91}.
Electrocardiologists have come to excel at visually interpreting
the detailed form of the ECG wave pattern and have
become adept at differential diagnoses.

Readily recognizable features of the ECG wave pattern are designated
by the letters P-QRS-T; the wave itself is often referred to as the
{\em QRS complex}.
Aside from the significance of various features of the QRS complex,
the timing of the {\em sequence} of QRS complexes over tens,
hundreds, and thousands of heartbeats is also significant.
These inter-complex times are readily measured by recording the
occurrences of the peaks of the large R waves which are, perhaps, the
most distinctive feature of the normal ECG.

In this Chapter we focus on various measures of the fluctuations of
this sequence of interbeat intervals and how such fluctuations can be
used to assess the presence or likelihood of cardiovascular disease
\cite{kitney82}.
This approach has come to be called {\em heart rate variability}
(HRV) analysis \cite{malik96,hon65} even when it is the time
{\em intervals} whose fluctuations are studied (heart rate has units of
inverse time rather than time).
HRV analysis serves as a marker for cardiovascular disease because
cardiac dysfunction is often manifested by systematic changes in the
variability of the RR-interval sequence relative to that of normal
controls \cite{levick91,malik96,kleiger87,bassingthwaighte94b}.
A whole host of HRV measures, some scale-dependent and others
scale-independent, have been developed and examined over the years in
an effort to develop readily available, inexpensive, and noninvasive
measures of cardiovascular function.

We examine sixteen HRV measures and their suitability for correctly
classifying ECG records of various lengths as normal or revealing the
presence of cardiac dysfunction.
Particular attention is devoted to HRV measures that are useful for
discriminating congestive-heart-failure patients from normal
subjects.
Using receiver-operating-characteristic (ROC) analysis we demonstrate
that scale-dependent HRV measures ({\em e.g.}, wavelet and spectral
measures) are substantially superior to scale-independent measures
(such as wavelet and spectral fractal exponents) for discriminating
these two classes of data over a broad range of record lengths.
The wavelet-transform standard deviation at a scale near 32 heartbeat
intervals, and its spectral counterpart near 1/32 cycles/interval,
turn out to provide reliable results using ECG records just minutes
long.

A long-standing issue of importance in cardiac physiology is the
determination of whether the normal RR sequence arises from a chaotic
attractor or has an underlying stochastic origin
\cite{bassingthwaighte94b}.
We present a phase-space analysis in which {\em differences} between
adjacent RR intervals are embedded.
This has the salutary effect of removing most of the correlation in
the time series, which is well-known to be deleterious to the detection
of underlying deterministic dynamics.
We demonstrate that RR sequences, from normal subjects and from patients
with cardiac dysfunction alike, have
stochastic rather than deterministic
origins, in accord with our earlier conclusions
\cite{turcott93,turcott96}.

Finally we develop a mathematical point process that emulates the
human heartbeat time series for both normal subjects and heart-failure
patients.
Using simulations, we show that a jittered integrate-and-fire model
built around a fractal-Gaussian-noise kernel provides a realistic,
though not perfect, simulation of real heartbeat sequences.
A construct of this kind may well be useful in a number of venues,
including pacemaker excitation.

\section{Methods and Measures}
\label{sec:methandmeas}

\subsection{The Heartbeat Sequence as a Point Process}

The statistical behavior of the sequence of heartbeats can be studied
by replacing the complex waveform of an individual heartbeat recorded
in the ECG (an entire QRS-complex) with the time of occurrence of the
contraction (the time of the peak of the R phase), which is a single
number \cite{turcott96,deboer84}.
In mathematical terms, the heartbeat sequence is then modeled as an
unmarked point process.
This simplification greatly reduces the computational complexity of
the problem and permits us to use the substantial methodology that
exists for point processes \cite{cox66,lewis72,saleh78}.

The occurrence of a contraction at time $t_i$ is therefore simply
represented by an impulse $\delta(t-t_i)$ at that time, where
$\delta$ is the Dirac delta function, so that the sequence of
heartbeats is represented by
\begin{equation}
h(t) =
\sum_i\delta(t-t_i).
\end{equation} A realization of a point
process is specified by the set of occurrence times $\{t_i\}$ of the
events.
A single realization of the data is often all that is available to
the observer so that the identification of the point process, and the
elucidation of the mechanisms that underlie it, must be gleaned from
this one realization.

One way in which the information in an experimental point process
can be made more digestible is to reduce the data into a statistic
that emphasizes a particular aspect of the data (at the expense of
other features).
These statistics fall into two broad classes which derive from the
sequence of interevent intervals and the sequence of counts, as
illustrated in Fig.~1 \cite{cox66,teich85}.

Figure~1 illustrates how an electrocardiogram may be analyzed to
obtain the sequence of interbeat intervals as well as the sequence of
counts.
Fig.~1(a) illustrates an ECG (sequence of QRS complexes) recorded
from a patient.
The R waves are schematically represented by a sequence of vertical
lines, as shown in Fig.~1(b).
The time between the first two R waves is $\tau_1$, the first RR (or
interbeat) interval, as indicated by the horizontal arrows in this
figure.
The time between the second and third R waves is $\tau_2$, and so
forth.
In Fig.~1(c), the time axis is divided into equally spaced,
contiguous time windows, each of duration $T$ seconds, and the
(integer) number of R waves that fall in the $i$th window is counted
and denoted $N_i$.
This sequence $\{N_i\}$ forms a discrete-time random counting process
of nonnegative integers.
Varying the duration $T$ yields a family of sequences $\{N_i\}(T)$.
The RR intervals $\{\tau_i\}$ themselves also form a sequence of
positive real-valued random numbers, which is shown schematically in
Fig.~1(d).
Here the abscissa is the interval number, which is not a
simple function of time.

In this section we examine several statistical measures (including
some that are novel) to characterize these stochastic processes; the
development is assisted by an understanding of point processes.

\subsubsection{Conventional Point Processes}

The homogeneous Poisson point process, perhaps the simplest of all
stochastic point processes, is described by a single parameter, the
rate $\lambda$. This point process is memoryless: the occurrence of
an event at any time $t_0$ is independent of the presence (or
absence) of events at other times $t \neq t_0$.
Because of this property, both the intervals $\{\tau_i\}$ and counts
$\{N_i\}$ form sequences of independent, identically distributed
random variables.
The homogeneous Poisson point process is therefore completely
characterized by the interevent-interval distribution (also referred
to as the interbeat-interval histogram), which is exponential, or the
event-number distribution (also referred to as the counting
distribution), which is Poisson, together with the property of being
independent.
This process serves as a benchmark against which other point
processes are measured; it therefore plays the role that the white
Gaussian process enjoys in the realm of continuous-time stochastic
processes.

A related point process is the nonparalyzable
fixed-dead-time-modified Poisson point process, a close cousin of the
homogeneous Poisson point process that differs only by the imposition
of a dead-time (refractory) interval after the occurrence of each
event, during which other events are prohibited from occurring
\cite{cox66,prucnal83}.
Another cousin is the gamma-$r$ renewal process which, for integer
$r$, is generated from an homogeneous Poisson point process by
permitting every $r$th event to survive while deleting all
intermediate events \cite{cox66,teich97}.
Both the dead-time-modified Poisson point process and the gamma-$r$
renewal process require two parameters for their description.

Some point processes exhibit no dependencies among their interevent
intervals at the outset, in which case the sequence of interevent
intervals forms a sequence of identically distributed random variables
and the point process is completely specified by its interevent-interval
histogram, i.e., its first-order statistic.
Such a process is called a renewal process \cite{cox66}, a definition
motivated by the replacement of failed parts, each replacement of
which forms a renewal of the point process.
Both examples of point processes presented above belong to the class
of renewal point processes.

The interevent-interval histogram is, perhaps, the most commonly used of all
statistical measures of point processes in the life sciences. The
interevent-interval histogram estimates the interevent-interval probability
density function $p_{\tau}(\tau)$ by computing the relative frequency of
occurrence of interevent intervals as a function of interval size. Its
construction involves the loss of interval ordering, and therefore of
information about dependencies among intervals; a reordering of the
sequence does not alter
the interevent-interval histogram since the order plays no role in the
relative frequency of occurrence.

The interevent-interval probability density function for the
homogeneous Poisson point process assumes the exponential form
\begin{equation} p_{\tau}(\tau)=\lambda \exp(-\lambda \tau)
\end{equation} where $\lambda$ is the mean number of events per unit
time.
The interevent-interval mean and variance are readily calculated to
be
${\rm E}[\tau]=\int_0^{\infty}\tau p_{\tau}(\tau)d\tau=1/\lambda$
and
${\rm Var}(\tau)= {\rm E}[\tau^2] - {\rm E}^2[\tau]=1/\lambda^2$,
respectively, where ${\rm E}[\cdot]$ represents expectation
over the quantity inside the brackets.
The interevent-interval probability density function for the
dead-time-modified Poisson point process exhibits the same
exponential form as for the homogeneous Poisson point process, but is
truncated at short interevent intervals as a result of the dead time
\cite{cox66}:
\begin{equation}
p_{\tau}(\tau)=\left\{ \begin{array}{ll} 0 & \mbox{$\tau<\tau_d$}
\\
\lambda \exp[-\lambda(\tau-\tau_d)] & \mbox{$\tau \ge \tau_d$}
\end{array}
\right.
\end{equation} Here $\tau_d$ is the dead time and $\lambda$ is the
rate of the process before dead time is imposed.

If a process is nonrenewal, so that dependencies exist among its
interevent intervals, then the interevent-interval histogram does not
completely characterize the process \cite{teich85}. In this case, measures that
reveal the nature of the dependencies provide information that is
complementary to that contained in the interevent-interval histogram.
The heartbeat time series is such a nonrenewal process.

\subsubsection{Fractal and Fractal-Rate Point Processes}
\label{sssec:ffrpp}

The complete characterization of a stochastic process involves a
description of all possible joint probabilities of the various events
occurring in the process.
Different statistics provide complementary views of the process; no
single statistic can in general describe a stochastic process
completely.
Fractal stochastic processes exhibit scaling in their statistics.
Such scaling leads naturally to power-law behavior, as demonstrated
in the following.
Consider a statistic $w$, such  as the Allan factor for long counting
times (see Sec.~\ref{sssec:afdef}), which depends continuously
on the scale $x$ over which measurements are taken
\cite{lowen95,thurner97}.
Suppose changing the scale by any factor $a$ effectively scales the
statistic by some other factor $g(a)$, related to the factor but
independent of the original scale:
\begin{equation} w(ax) = g(a)w(x).
\end{equation}
The only nontrivial solution of this scaling equation, for real
functions and arguments, that is independent of $a$ and $x$ is
\begin{equation} w(x) = bg(x) \hspace{7mm} \mbox{with} \hspace{7mm}
g(x) = x^{c}
\label{eq:scale0}
\end{equation} for some constants $b$ and $c$
\cite{lowen95,thurner97,rudin76}.
Thus statistics with power-law forms are closely related to the
concept of a fractal \cite{mandelbrot83,feder88,liebovitch98}.
The particular case of fixed $a$ admits a more general solution
\cite{shlesinger91}:
\begin{equation} g(x;a)= x^{c}\cos[2\pi\ln(x)/\ln(a)].
\end{equation}

Consider once again, for example, the interevent-interval histogram. This
statistic highlights the behavior of the times between adjacent events, but
reveals none of the information contained in the relationships among these
times, such as correlation between adjacent time intervals.
If the interevent-interval probability density function follows the
form of Eq.~(\ref{eq:scale0}) so that $p(\tau) \sim \tau^{c}$ over
a certain range of $\tau$ where $c< -1$, the process is known as a
fractal renewal point process \cite{thurner97,mandelbrot83}, a form
of fractal stochastic process.

A number of statistics may be used to describe a fractal stochastic
point process, and each statistic which scales will in general have a
different scaling exponent $c$.
Each of these exponents can be simply related to a more general
parameter $\alpha$, the fractal exponent, where the exact relation
between these two exponents will depend upon the statistic in
question.
For example, the exponent $c$ of the interevent-interval probability
density function defined above is related to the fractal exponent
$\alpha$ by $c = -(1 + \alpha)$.
As the fractal exponent is a constant that describes the overall
scaling behavior of a statistic, it does not depend on the particular
scale and is therefore {\em scale independent}.
Scale-independent measures are discussed in
subsections~\ref{ssec:scaleind} and~\ref{sssec:sivssd}.

Sample functions of the fractal renewal point
process are true fractals; the expected value of their generalized
dimensions assumes
a nonintegral value between the topological dimension (zero) and the
Euclidean dimension (unity) \cite{mandelbrot83}.

The sequence of unitary events observed in many biological and
physical systems, such as the heartbeat sequence, do not exhibit
power-law-distributed
interevent-interval histograms but nevertheless exhibit scaling in
other statistics.
These processes therefore have integral generalized dimensions and
are consequently not true fractals.
They may nevertheless be endowed with rate functions that are either
fractals or their increments: fractal
Brownian motion, fractal Gaussian noise, or other
related processes.
Therefore, such point processes are more properly termed fractal-{\em
rate} stochastic point processes \cite{thurner97}.
It can be shown by surrogate data methods, e.g., shuffling the order
of the intervals (see Sec.~\ref{sssec:surrogate}), that it is
the ordering and not the relative interval sizes that distinguish
these point processes \cite{turcott96}.

\subsection{Standard Frequency-Domain Measures}
\label{ssec:stdfreqm}

A number of HRV measures have been used as standards in cardiology,
both for purposes of physiological interpretation and for clinical
diagnostic applications \cite{malik96}.
We briefly describe some of the more commonly used measures that we
include in this chapter for comparison with several novel measures
that have been recently developed.

Fourier transform techniques provide a method for quantifying the
correlation properties of a stochastic process through spectral
analysis.
Two definitions of power spectral density  have been used in the
analysis of HRV \cite{deboer84}.
A {\em rate}-based power spectral density $S_\lambda(f)$ is obtained by
deriving an underlying random continuous process $\lambda(t)$, the heart
rate, based on a transformation of the observed RR interbeat
intervals.
The power spectral density of this random process is well defined,
and standard techniques may be used for estimating the power spectral
density  from a single observation of $\lambda(t)$.
An advantage of this technique is that the power spectral density
thus calculated has temporal frequency as the independent variable,
so that spectral components can be interpreted in terms of underlying
physiological processes with known timescales.
The power spectral density itself is usually expressed in units of
sec$^{-1}$.
However, the choice of how to calculate $\lambda(t)$, the underlying rate
function, may influence the calculated power spectral density.

The second spectral measure that is more widely used is an
{\em interval}-based power spectral density $S_\tau(f)$ that is
directly calculated from measured RR interbeat intervals without
transformation \cite{deboer84}.
In this case the intervals are treated as discrete-index samples of
an underlying random process, and there is no intermediate
calculation of an underlying rate function.
The power spectral density in this case has cycles/interval as
the independent variable, and therefore has units of sec$^2/$interval.

The two types of power spectral densities are easily confused and
care must be taken in their interpretation.
For example, one could mistakenly interpret the abscissa of an
interval-based power spectral density plot as being equivalent to
temporal frequency (e.g., cycles/sec).
While this is generally incorrect, for point processes whose
interevent-interval coefficient of variation is relatively small
\cite{deboer84}, the interval-based and rate-based power spectral
density plots can be made approximately equivalent by converting the
interval-based frequency $f_{\rm int}$ (in cycles/interval) to
the time-based frequency $f_{\rm time}$ (in cycles/sec) using
\begin{equation}
f_{\rm time}=f_{\rm int}/{\rm E}[\tau].
\label{eq:freqconv}
\end{equation}
For typical interbeat-interval sequences, the coefficient of
variation is indeed relatively small and this conversion can be
carried out without the introduction of significant error
\cite{deboer84}.
In the remainder of this Chapter we work principally with the
interval-based power-spectral density. We use
the notation $f \equiv f_{\rm int}$ for the interval-based frequency
(cycles/interval) and retain the notation $f_{\rm time}$ for
temporal frequency (cycles/sec).

We make use of a non-parametric technique for estimating the spectral
density.
A simple reliable method for estimating the power spectral density of
a process from a set of discrete samples $\{ \tau_i\}$ is to
calculate the averaged periodogram
\cite{papoulis91,oppenheim89,press92}.
The data is first divided into $K$ non-overlapping blocks of $L$
samples. After the optional use of a Hanning window,
the discrete Fourier transform of each block is calculated and
squared.
The results are then averaged to form the estimate
\begin{equation}
\widehat{S_{\tau}}(f)\equiv {{1}\over
{K}}\sum_{k=1}^K|\tilde\tau_k(f)|^2.
\label{eq:avgpgdef}
\end{equation}
Here $\tilde\tau_k(f)$ is the discrete Fourier transform of the $k$th
block of data and the hat explicitly indicates that we are dealing
with an {\em estimate} of $S_{\tau}(f)$, which is called an averaged
periodogram.

The periodogram covers a broad range of frequencies which can be
divided into bands that are relevant
to the presence of various
cardiac pathologies.
The power within a band is calculated by integrating the power
spectral density over the associated frequency range.
Some commonly used measures in HRV are \cite{malik96}:

\noindent {\bf VLF.} The power in the very-low-frequency range:
0.003--0.04 cycles/interval.
Physiological correlates of the VLF band have not been specifically
identified \cite{malik96}.

\noindent {\bf LF.} The power in the low-frequency range: 0.04--0.15
cycles/interval.
The LF band may reflect both sympathetic and vagal activity but its
interpretation is controversial \cite{malik96}.

\noindent {\bf HF.} The power in the high-frequency range: 0.15--0.4
cycles/interval.
Efferent vagal activity is a major contributor to the HF band
\cite{akselrod81,pomeranz85,malliani91}.

\noindent {\bf LF/HF.} The ratio of the low-frequency-range power to
that in the high-frequency range.
This ratio may mirror either sympatho-vagal balance or reflect
sympathetic modulations \cite{malik96}.

\subsection{Standard Time-Domain Measures}

We consider three time-domain measures commonly used in HRV analysis.
The first and last are highly correlated with each other inasmuch as
they estimate the high-frequency variations in the heart rate
\cite{malik96}.
They are:

\noindent {\bf pNN50.} The relative {\bf p}roportion of successive
{\bf NN} intervals (normal-to-normal intervals, i.e., all intervals
between adjacent QRS complexes resulting from sinus node
depolarizations \cite{malik96}) with interval differences greater
than {\bf 50} ms.

\noindent {\bf SDANN.} The {\bf S}tandard {\bf D}eviation of the
{\bf A}verage {\bf NN} interval calculated in five-minute segments.
It is often calculated over a 24-hour period.
This measure estimates fluctuations over frequencies smaller than
$0.003$ cycles/sec.

\noindent {\bf SDNN \boldmath ($\sigma_{\rm int}$).}
The {\bf S}tandard {\bf D}eviation of the {\bf NN} interval set
$\{\tau_i\}$ specified in units of seconds.
This measure is one of the more venerable among the many
scale-dependent measures that have long been used for HRV analysis
\cite{malik96,kleiger87,wolf78,saul88}.

\subsection{Other Standard Measures}

There are several other well-known measures that have been considered
for HRV analysis.
For completeness, we briefly mention two of them here: the
event-number histogram and the Fano factor \cite{turcott93,turcott96}.
Just as the interevent-interval histogram provides an estimate of the
probability density function of interevent-interval magnitude, the
event-number histogram provides an estimate of the probability mass
function of the number of events.
Construction of the event-number histogram, like the
interevent-interval histogram, involves loss of information, in this
case the ordering of the counts.  However, whereas the time scale of
information contained in the interevent-interval histogram is the
mean interevent interval, which is intrinsic to
the process under consideration, the event-number histogram reflects
behavior occurring on the adjustable time scale of the counting
window $T$.
The Fano factor, which is the variance of the number of events in a
specified counting time $T$ divided by the mean number of events in
that counting time, is a measure of correlation over different time
scales $T$.
This measure is sometimes called the index of dispersion of counts
\cite{cox80}. In terms of the sequence of counts illustrated in Fig.~1,
the Fano factor is simply the variance of $\{ N_{i} \}$ divided by the
mean of $\{ N_{i}\}$.

\subsection{Novel Scale-Dependent Measures}

The previous standard measures are all well-established scale-dependent
measures.
We now  describe a set of recently devised scale-dependent measures whose
performance we evaluate.
Throughout this chapter, when referring to intervals, we denote the
fixed scale as $m$; when referring to time, we employ $T$.

\subsubsection[Allan Factor {$[A(T)]$}]{Allan Factor \boldmath $[A(T)]$}
\label{sssec:afdef}

In this section we present a measure we first defined in 1996
\cite{lowen96b} and called the Allan factor. We quickly found
that this quantity
was a useful measure of HRV \cite{turcott96}. The Allan factor is the
ratio of the event-number Allan variance to twice the mean:
\begin{equation}
A(T) \equiv \frac{\displaystyle {{\rm E} \left\{\left[N_{i+1}(T) -
N_{i}(T)\right]^2\right\}}} {\displaystyle {2 {\rm E}\{N_{i+1}(T)\}}}.
\label{eq:afdef}
\end{equation}
The Allan variance, as opposed to the ordinary variance, is defined
in terms of the variability of {\em successive} counts
\cite{thurner97,allan66,barnes66}.
As such, it is a measure based on the Haar wavelet.
The Allan variance was first introduced in connection with the
stability of atomic-based clocks \cite{allan66}.
Because the Allan factor functions as a derivative, it has the
salutary effect of mitigating against linear nonstationarities.

The Allan factor of a point process generally varies as a function of
the counting time $T$; the exception is the homogeneous Poisson point
process.
For a homogeneous Poisson point process, $A(T)=1$ for any counting
time $T$.
Any deviation from unity in the value of $A(T)$ therefore indicates
that the point process in question is not Poisson in nature.
An excess above unity reveals that a sequence is less ordered than a
homogeneous Poisson point process, while values below unity signify
sequences which are more ordered.
For a point process without overlapping events the Allan factor
approaches unity as $T$ approaches zero.

A more complex wavelet Allan factor can be constructed to eliminate
polynomial trends \cite{teich96c,teich96a,abry96}.
The Allan variance, ${\rm E}[(N_{i+1}-N_{i})^{2}]$ may be recast as the
variance of the integral of the point process under study multiplied
by the following function:
\begin{equation}
\psi_{\rm Haar}(t) =
\left\{ \begin{array}{ll} -1 & \mbox{ for $-T < t < 0$,}\\ +1 &
\mbox{ for $0 < t < T$,}\\ 0 & \mbox{ otherwise.}
\end{array} \right.
\label{eq:haar}
\end{equation} Equation~(\ref{eq:haar}) defines a scaled wavelet
function, specifically the Haar wavelet. This can be generalized to
any admissible wavelet $\psi(t)$; when suitably normalized the
result is a wavelet Allan factor
\cite{teich96a,heneghan96}.

\subsubsection[Wavelet-Transform Standard Deviation
{[$\sigma_{\rm wav}(m)$]}]{Wavelet-Transform Standard Deviation
\boldmath [$\sigma_{\rm wav}(m)$]}
\label{sssec:wtsddef}

Wavelet analysis has proved to be a useful technique for analyzing
signals at multiple
scales~\cite{mallat89,meyer90,daubechies92,aldroubi96,akay97a,akay97b}.
It permits the time and frequency characteristics of a signal to be
simultaneously examined, and has the advantage of naturally
removing polynomial nonstationarities~\cite{teich96a,abry96,arneodo88}.
The Allan factor served in this capacity for the
counting process $\{N_i\}$, as discussed above.
Wavelets similarly find use in the analysis of RR-interval series.
They are attractive because they mitigate against the
nonstationarities and slow variations inherent in the
interbeat-interval sequence.
These arise, in part, from the changing activity level of the subject
during the course of a 24-hour period.

Wavelet analysis simultaneously gives rise to both scale-dependent
and scale-independent measures \cite{thurner98}, affording the
experimenter an opportunity to compare the two approaches.
In this latter capacity wavelet analysis provides an estimate of the
wavelet-transform fractal (scaling) exponent $\alpha_W$
\cite{thurner98,herz2}, as discussed in the context of HRV in
subsections~\ref{sssec:wtple} and~\ref{sssec:sivssd}.
As a result of these salutary properties we devote particular
attention to the wavelet analysis of HRV in this Chapter.

A dyadic discrete wavelet transform for the RR-interval sequence
$\{ \tau_i \}$ may be defined as \cite{daubechies92,aldroubi96,akay97a}
\begin{equation}
W_{m,n}(m) = {{1}\over{\sqrt{m}}} \sum_{i=0}^{L-1} \tau_{i}
\psi (i/m-n).
\label{eq:wtdef}
\end{equation}
The quantity $\psi$ is the wavelet basis function, and $L$ is the
number of RR intervals in the set $\{\tau_i\}$.
The scale $m$ is related to the scale index $j$ by $m=2^j$.
Both $j$ and the translation variable $n$ are nonnegative integers.
The term dyadic refers to the use of scales that are integer powers
of 2.
This is an arbitrary choice; the wavelet transform could be
calculated at arbitrary scale values, although the dyadic scale
enjoys a number of convenient mathematical properties
\cite{daubechies92,aldroubi96}.

The dyadic discrete wavelet transform calculated according to
this prescription generates a  three-dimensional
space from a two-dimensional signal graph.
One axis is time or, in our case, the RR-interval number $i$; the
second axis is the scale $m$; and the third axis is the strength of
the wavelet component.
Pictorially speaking, the transform gives rise to a landscape whose
longitude and latitude are RR-interval number and scale of
observation, while the altitude is the value of the discrete wavelet
transform at the interval $i$ and the scale $m$.

Figure~2 provides an example of such a wavelet transform, where $\psi
(x)$ is the simple Haar wavelet.
Figure 2(a) illustrates the original wavelet, a function that is by
definition $\psi(x) = 1$ for $x$ between 0 and 0.5; $\psi(x) = -1$
for $x$ between 0.5 and 1; and $\psi(x) = 0$ elsewhere.
Figure 2(b) illustrates the wavelet scaled by the factor $m=16$,
which causes it to last for 16 samples rather than 1; and delayed by
a factor of $n=3$ times the length of the wavelet, so that it begins
at $x=nm=48$.
Figure 2(c) shows a sequence of interbeat-interval values multiplied
by the scaled and shifted wavelet [the summand in
Eq.~(\ref{eq:wtdef})].
The abscissa is labeled $i$ rather than $x$ to indicate that we have
a discrete-time process comprised of the sequence
$\{\tau_i\}$.
In this particular example, only values of $\tau_i$ between $i=48$
and 63 survive.
Adding them (with the appropriate sign) provides the wavelet
transform beginning at interval number $i=48$ at a scale of
$m=16$.

For the Haar wavelet
the calculation of the wavelet transform is therefore tantamount
to adding the eight RR intervals between intervals 48 and 55
inclusive, and then subtracting the eight subsequent RR intervals
between intervals 56 and 63 inclusive, as illustrated in Fig.~2(c).
Moving this window across interval number allows us to see how the
wavelet transform evolves with interval number, whereas varying the
scale of the window permits this variation to be observed over a
range of resolutions, from fine to coarse (smaller scales allow the
observation of more rapid variations, i.e. higher frequencies).

A simple measure that can be devised from the wavelet transformation is the
standard deviation of the wavelet transform as a function of scale
\cite{thurner98,herz2,teich98,heneghan98}:
\begin{equation}
\sigma_{\rm wav}(m) =
\left[{\rm E}\left\{ | W_{m,n}(m) -
{\rm E}[W_{m,n}(m)] |^2 \right\}\right]^{1/2}
\end{equation}
where the expectation is taken over the process of RR intervals, and
is independent of $n$.
It is readily shown that ${\rm E}[W_{m,n}(m)]=0$ for all values of
$m$ so that a simplified form for the wavelet-transform standard
deviation emerges:
\begin{equation}
\sigma_{\rm wav}(m) =
\left\{{\rm E}\left[| W_{m,n}(m) |^2\right] \right\}^{1/2}
\label{eq:wtsddef}
\end{equation}
This quantity has recently been shown to be quite valuable for HRV
analysis \cite{thurner98,herz2,teich98,heneghan98,ashkenazy98}.
The special case obtained by using the Haar-wavelet basis and
evaluating Eq.~(\ref{eq:wtsddef}) at $m=1$ yields the standard
deviation of the difference between pairs of consecutive interbeat
intervals.
This special case is therefore identical to the well-known HRV
measure referred to as {\bf RMSSD} \cite{malik96}, an abbreviation for
{\bf R}oot-{\bf M}ean-{\bf S}quare of {\bf S}uccessive-interval
{\bf D}ifferences.

Figure~3 provides an example in which the discrete wavelet
transform is calculated using an RR-interval data set. In
Fig.~3(a), the original RR interbeat-interval series is shown, while Fig.~3(b)
shows the dyadic discrete wavelet transform at three different scales as
a function of RR-interval number. It is important and interesting
to note that the trends and baseline variations
present in the original time series have been removed by the
transform. As illustrated in Fig.~3(b), the wavelet-transform
standard deviation
$\sigma_{\rm wav}$ typically increases with the
scale
$m$. When plotted versus scale, this quantity provides information
about the
behavior of the signal at all scales.  In Sec.~\ref{sec:discrim}
we show how this measure can be effectively used to separate
heart-failure patients from normal normal subjects.

\subsubsection[Relationship of Wavelet {[$\sigma_{\rm wav}(m)$]} and
Spectral Measures {[$S_{\tau}(f)$]}]{Relationship of Wavelet
\boldmath [$\sigma_{\rm wav}(m)$] and Spectral Measures
[$S_{\tau}(f)$]}
\label{sssec:wtandspec}

Is there a spectral measure equivalent to the wavelet-transform
standard deviation?
We proceed to show that the wavelet-transform standard
deviation $\sigma_{\rm wav}(m)$ and the
interval-based power spectral density $S_{\tau}(f)$
are isomorphic \cite{heneghan98}, so that the answer is yes
under conditions of stationarity.
Though their equivalence is most easily
analyzed in the continuous
domain, the results are readily translated to the
discrete domain by interpreting the discrete
wavelet transform as a discretized version of a continuous wavelet transform.

The
continuous wavelet transform of a signal $\tau(t)$ is defined as
\begin{equation}
W_{\tau}(s,r)={{1}\over{\sqrt{s}}}\int_{-\infty}^{\infty}
\tau(t)\psi^*\left({{t-r}\over{s}}\right)dt
\end{equation}
where $s$ and $r$ are continuous-valued scale and translation
parameters respectively, $\psi$ is a wavelet basis function, and $*$
denotes complex conjugation.
Since ${\rm E}[W_{\tau}]=0$, the variance of $W_{\tau}$ at scale $s$ is
\begin{equation}
D(s) =\sigma_{\rm wav}^2(s)={\rm E}\left[|W_{\tau}(s,r)|^2\right],
\end{equation}
which can be written explicitly as
\begin{equation}
D(s)  ={\rm E}\left[{{1}\over{\sqrt{s}}}
\int_{-\infty}^\infty\tau(t)\psi^*\left({{t-r}\over{s}}\right)dt
{{1}\over{\sqrt{s}}}
\int_{-\infty}^\infty\tau^*(t')\psi\left({{t'-r}\over{s}}\right)dt'
\right].
\end{equation}
For a wide-sense stationary signal the variance can be written as
\begin{equation}
D(s)={{1}\over{s}}\int_{-\infty}^\infty\int_{-\infty}^\infty
R(t-t')\psi^*\left({{t-r}\over{s}}\right)
\psi\left({{t'-r}\over{s}}\right)dtdt'
\end{equation}
where $R$ is the autocorrelation function of $\tau$. Routine algebraic
manipulation then leads to
\begin{equation}
D(s)=s\int_{-\infty}^{\infty}R(sy)W_{\psi}(1,y)dy
\end{equation}
or, alternatively,
\begin{equation}
D(s)=s\int_{f=-\infty}^{\infty}S_{\tau}(f)\left
[\int_{y=-\infty}^{\infty}W_{\psi}(1,y)\exp(j2\pi fsy)dy
\right]df
\label{eq:cwtvar}
\end{equation}
where $W_{\psi}(1,y)$ is the wavelet transform of the wavelet itself
(termed the wavelet kernel), and $S_{\tau}(f)$ is the power spectral
density of the signal.

For the dyadic discrete wavelet transform that we have used,
Eq.~(\ref{eq:cwtvar}) becomes
\begin{eqnarray}
D(m)=\sigma_{\rm wav}^2(m) & = &
m\int_{f=-\infty}^{\infty}S_{\tau}(f)\left[
\int_{y=-\infty}^{\infty}W_{\psi}(1,y)\exp(j2\pi fmy)dy
\right]df \nonumber \\
& = & m \int^\infty_{-\infty}S_{\tau}(f)H(mf)df.
\label{eq:dwtvar}
\end{eqnarray}
We conclude that for stationary signals the interval-based power
spectral density $S_{\tau}(f)$ is directly related to the
wavelet-transform standard deviation $\sigma_{\rm wav}(m)$
through an integral transform.
This important result has a simple interpretation: the factor in
square brackets in Eq.~(\ref{eq:dwtvar}) represents a bandpass filter
$H(mf)$ that
only passes spectral components in a bandwidth surrounding the
frequency $f_m$ that corresponds to the scale $m$.
This is because the Fourier transform of a wavelet kernel is
constrained to be bandpass in nature.
For a discrete-index sequence, the sampling ``time'' can be
arbitrarily set to unity so that a frequency $f_m$ corresponds to
$1/m$.
We conclude that information obtained from a $D(m)$-based statistic
is also accessible through interval-based power spectral
density measures.
In Sec.~\ref{sec:discrim} we explicitly show that the two measures
are comparable in their abilities to discriminate heart-failure
patients from normal subjects.

\subsubsection[Detrended Fluctuation Analysis {[DFA($m$)]}]{Detrended
Fluctuation Analysis \boldmath {\bf [DFA($m$)]}}
\label{sssec:dfadef}

Detrended fluctuation analysis (DFA)
was originally proposed as a technique for quantifying the nature of
long-range correlations in a time series \cite{peng95,peng96,ho97}. As
implied by its name, it was conceived as a method for
detrending variability in a sequence of events.
The DFA computation involves the calculation of the summed series
\begin{equation}
y(k)=\sum_{i=1}^k\{\tau_i-{\rm E}[\tau]\}
\label{eq:sumser}
\end{equation}
where $y(k)$ is the $k$th value of the summed series and
${\rm E}[\tau]$ denotes the average over the set $\{\tau_i\}$.
The summed series is then divided into segments of length $m$ and a
least-squares fit is performed on each of the data segments,
providing the trends for the individual segments.
Detrending is carried out by subtracting the local trend $y_m(k)$ in
each segment.
The root-mean-square fluctuation of the resulting series is then
\begin{equation}
F(m)=\left\{{{1}\over{L}}\sum_{k=1}^L[y(k)-y_m(k)]^2\right\}^{1/2}.
\label{eq:dfadef}
\end{equation}
The functional dependence of $F(m)$ is obtained by evaluations over
all segment sizes $m$.

Although detrended fluctuation analysis was originally proposed as a
method for estimating the scale-independent fractal exponent of a
time series \cite{peng95}, as discussed in
Sec.~\ref{sssec:dfaple}, we consider its merits as a scale-{\em
dependent} measure.
As will be demonstrated in Sec.~\ref{sec:discrim}, a plot of
$F(m)$ versus $m$ reveals a window of separation between
congestive-heart-failure patients and normal subjects over a limited
range of scales, much as that provided by the other scale-dependent
measures discussed in this section.
Because DFA is an {\em ad hoc} measure that involves
nonlinear computations it is difficult to relate it to other
scale-dependent measures in the spirit of Eq.~(\ref{eq:dwtvar}).
Furthermore, as will become clear in
Sec.~\ref{sssec:comptimes}, relative to other measures
DFA is highly time intensive from a
computational point-of-view.

\subsection{Scale-Independent Measures}
\label{ssec:scaleind}

Scale-independent measures are designed to estimate fractal exponents
that characterize scaling behavior in one or more statistics of a
sequence of events, as discussed in Sec.~\ref{sssec:ffrpp}.
The canonical example of a scale-independent measure in HRV is the fractal
exponent $\alpha_S$ of the interbeat-interval power spectrum,
associated with the decreasing power-law form of the spectrum at
sufficiently low frequencies $f$: $S_{\tau}(f) \propto f^{-{\alpha_S}}$
\cite{malik96,saul88,kobayashi82}.
Other scale-independent measures have been examined by us
\cite{turcott93,turcott96,lowen95,thurner97,thurner98} and by others
\cite{peng95,peng93,amaral98,bassingthwaighte94b} in connection with
HRV analysis.
For exponent values encountered in HRV, and infinite data
length, all measures should
in principle lead to a unique fractal exponent.  In practice, however,
finite data length and other factors
introduce bias and variance, so that different measures give
rise to different results.
The performance of scale-independent
measures has been compared with that of scale-dependent
measures for assessing cardiac dysfunction \cite{thurner98,herz2}.

\subsubsection[Detrended-Fluctuation-Analysis Power-Law Exponent
{($\alpha_{D}$)}]{Detrended-Fluctuation-Analysis Power-Law Exponent
\boldmath ($\alpha_{D}$)}
\label{sssec:dfaple}

The DFA technique, and its use as a scale-dependent measure, has been
described in Sec.~\ref{sssec:dfadef}.
A number of recent studies \cite{peng95,ho97,amaral98} have
considered the extraction of power-law exponents from DFA and their
use in HRV.
As originally proposed \cite{peng95}, $\log[F(m)]$ is plotted against
$\log(m)$ and scaling exponents are obtained by fitting straight
lines to sections of the resulting curve -- the exponents are simply
the slopes of the linearly fitted segments on this doubly logarithmic
plot.
The relationship between the scaling exponents has been proposed as a
means of differentiating normal from pathological subjects
\cite{peng95,peng93,amaral98}.

\subsubsection[Wavelet-Transform Power-Law Exponent
{($\alpha_W$)}]{Wavelet-Transform Power-Law Exponent \boldmath
($\alpha_W$)}
\label{sssec:wtple}

The use of the wavelet transform as a scale-dependent measure was
considered in Sec.~\ref{sssec:wtsddef}.
It was pointed out that a scale-independent measure also emerges from
the wavelet-transform standard deviation.
The wavelet-transform fractal exponent $\alpha_W$
is estimated directly from the wavelet
transform as twice the slope of the curve $\log[\sigma_{\rm wav}(m)]$
versus $\log(m)$, measured at large values of $m$ \cite{thurner98}.
The factor of two is present because the fractal exponent is related
to variance rather than to standard deviation.

\subsubsection[Periodogram Power-Law Exponent
{($\alpha_S$)}]{Periodogram Power-Law Exponent \boldmath
$(\alpha_S)$}

The description of the periodogram as a scale-dependent measure was
provided in Sec.~\ref{ssec:stdfreqm}.
The periodogram fractal exponent $\alpha_S$
\cite{thurner97,kobayashi82} is obtained as the
least-squares-fit slope of the spectrum when
plotted on doubly logarithmic coordinates.
The range of low frequencies over which the slope is estimated
stretches between $10/L$ and $100/L$ where $L$ is the length of the
data set \cite{thurner97}.

\subsubsection[Allan-Factor Power-Law Exponent
{($\alpha_A$)}]{Allan-Factor Power-Law Exponent \boldmath
($\alpha_A$)}

The use of the Allan factor as a scale-dependent measure was
considered in Sec.~\ref{sssec:afdef}.
The Allan factor fractal exponent $\alpha_A$
\cite{turcott96,thurner97} is obtained by determining the slope of
the best-fitting straight line, at large values of $T$, to the Allan
factor curve [Eq.~(\ref{eq:afdef})] plotted on doubly logarithmic coordinates.
Estimates of
$\alpha$ obtained from the Allan factor can range up to a value of three
\cite{lowen96b}. The use of wavelets more complex than the Haar enables an
increased range of fractal exponents to be accessed, at the cost of
a reduction in the range of counting time over which the wavelet Allan factor
varies as
$~T^{\alpha_A}$. In general, for a particular wavelet with regularity
(number of vanishing moments) $R$, fractal exponents $\alpha < 2R +
1$ can be reliably estimated \cite{teich96a,heneghan96}. For the Haar
basis, $R=1$ whereas all other wavelet bases have $R>1$.
A wavelet Allan factor making use of bases other than the Haar is
therefore required for fractal-rate stochastic point processes for
which $\alpha \ge 3$.
For processes with $\alpha < 3$, however, the Allan factor appears to
be the best choice \cite{teich96a,heneghan96}.

\subsubsection[Rescaled-Range-Analysis Power-Law Exponent
{($\alpha_R$)}]{Rescaled-Range-Analysis Power-Law Exponent \boldmath
($\alpha_R$)}

Rescaled range analysis \cite{mandelbrot83,hurst51,feller51,schepers92},
provides information about correlations
among blocks of interevent intervals. For a block of $k$ interevent
intervals, the difference between each interval and the mean
interevent interval is obtained and successively added to a
cumulative sum. The normalized range $R(k)$ is the difference
between the maximum and minimum values that the cumulative sum
attains, divided by the standard deviation of the interval size.
$R(k)$ is plotted against $k$. Information about the nature and the
degree of correlation in the process is obtained by fitting $R(k)$
to the function $k^H$, where
$H$ is the so-called Hurst exponent \cite{hurst51}. For $H > 0.5$
positive correlation exists among the intervals, whereas $H < 0.5$
indicates the presence of negative correlation;
$H = 0.5$ obtains for intervals with no correlation. Renewal
processes yield $H = 0.5$. For negatively correlated intervals, an
interval that is larger than the mean tends, on average, to be
preceded or followed by one smaller than the mean.

The Hurst exponent $H$ is generally assumed to be well suited to
processes that exhibit long-term correlation or have a large variance
\cite{mandelbrot83,hurst51,feller51,schepers92}, but there are limits
to its robustness since it exhibits large systematic errors and
highly variable estimates for some fractal sequences
\cite{bassingthwaighte94b,beran94,bassingthwaighte94a}.
Nevertheless, it provides a useful indication of correlation in a
sequence of events arising from the ordering of the interevent
intervals alone.

The exponent $\alpha_R$ is ambiguously related to the Hurst exponent $H$,
since some authors have used the quantity $H$ to index fractal
Gaussian noise whereas others have used the same value of $H$ to
index the integral of fractal Gaussian noise (which is fractional
Brownian motion). The relationship between the quantities is $\alpha_R
= 2H - 1$ for fractal Gaussian noise and $\alpha_R = 2H + 1$ for
fractal Brownian motion. In the context of this work, the former
relationship holds.

\subsection{Estimating the Performance of a Measure}

We have, to this point, outlined a variety of candidate measures for
use in HRV analysis.
The task now is to determine the relative value of these measures
from a clinical perspective.
We achieve this by turning to estimation theory \cite{vantrees68}.

A statistical measure obtained from a finite set of actual data is
characterized by an estimator. The fidelity with which the estimator
can approximate the true value of the measure is determined by
its bias and variance. The bias is the deviation of the expected
value of the estimator from its true underlying value
(assuming that this exists) whereas the variance
indicates the expected deviation from the the mean.
An ideal estimator has zero bias and zero variance, but this is not
achievable with a finite set of data. For any unbiased estimator the
Cram\'{e}r-Rao bound provides a
lower bound for the estimator variance; measures that
achieve the Cram\'{e}r-Rao bound are called efficient estimators.
The estimator bias and variance play a role in establishing
the overall statistical significance of conclusions based on the
value returned by the estimator.

\subsubsection[Statistical Significance: $p$, $d'$, $h$, and
$d$]{Statistical Significance: \boldmath $p$, $d'$, $h$, and $d$}
\label{sssec:statsig}

The concept of statistical significance extends the basic
properties of bias and variance \cite{hald52}.
It provides a probabilistic interpretation of how likely
it is that a particular value
of the estimator might occur by chance alone, arising from both random
fluctuations in the
data and the inherent properties of the estimator itself.

A frequently used standard of statistical significance
is the $p$-value, the calculation of which almost always implicitly assumes a
Gaussian-distributed dependent variable. A lower
$p$-value indicates greater statistical significance, and a measure is
said to be statistically significant to a value of $p_0$ when $p<p_0$.
The distributions
obtained from HRV measurements are generally not Gaussian, however,
so that the usual method for
estimating the $p$-value cannot be used with confidence.
Since other methods for estimating
the $p$-value
require more data than is available we do not consider this quantity
further.

Another often-used distribution-dependent standard is the
$d'$-value. It serves to indicate the degree of separation
between two distributions, and has been widely used in signal
detection theory
and psychophysics where the two distributions represent noise
and signal-plus-noise \cite{green66}. The most common definition of
$d'$ is the
difference in the means of two Gaussian distributions divided by
their common standard deviation. Two closely related
distribution-dependent cousins of $d'$ are the detection distance
$h$, defined as
the difference in the means of the two Gaussian distributions divided by
the square-root of the sum of their variances; and
the detection distance $d$, defined as the difference in the means
of the two
Gaussian distributions divided by the sum of their standard
deviations.
Larger values of $d', h$, and $d$ indicate
improved separation between
the two distributions and therefore reduced error in assigning
an outcome to one or the other of the
hypotheses.

Because HRV measures are intended to provide diagnostic information
in a clinical setting, and do not return Gaussian statistics,
the evaluation of their performance using
distribution-independent means is preferred.
Two techniques for achieving this,
positive and negative predictive values, and receiver
operator characteristic (ROC) analysis, are described below.
Neither requires knowledge of
the statistical distribution of the measured
quantities and both are useful.

\subsubsection{Positive and Negative Predictive Values}
\label{sssec:pvdef}

The performance of the various HRV measures discussed previously
can be effectively compared using
positive predictive values
and negative predictive
values,
the proportion of correct positive and negative identifications
respectively.
When there is no false positive (or negative) detection, the
predictive value is equal to unity and there is perfect assignment.
Furthermore, when the individual values of a measure for normal
subjects and patients do not overlap, the predictive value curves are
typically monotonic, either increasing or decreasing, with the
threshold.
A detailed discussion of positive and negative predictive values is
provided in Sec.~\ref{ssec:pvplots}.

\subsubsection{Receiver-Operating-Characteristic (ROC) Analysis}
\label{sssec:rocdef}

Receiver-operating-characteristic (ROC)
analysis \cite{vantrees68,green66,swets88} is an
objective and highly effective technique for
assessing the performance of a measure when it is used in binary
hypothesis
testing. This format provides that a data sample be assigned to one
of two hypotheses or classes (e.g., pathologic or normal)
depending on the value of some measured statistic relative to a
threshold value. The
efficacy of a measure is then judged on the basis of its sensitivity
(the proportion of pathologic patients correctly identified) and its
specificity (the proportion of normal subjects correctly
identified).
The ROC curve is a graphical presentation of sensitivity versus
$1-$specificity as a threshold parameter is swept.
Note that sensitivity and specificity relate to the status of the
patients (pathologic and normal) whereas predictive values relate
to the status of the identifications (positive and negative).

The area under the ROC curve serves as a well-established index of
diagnostic accuracy \cite{green66,swets88}; the maximum value of 1.0
corresponds to perfect assignment (unity sensitivity for all values
of specificity) whereas a value of 0.5 arises from
assignment to a class by pure chance (areas $<0.5$ arise when
the sense of comparison is reversed).
ROC analysis can be used to choose the best of a host of different
candidate diagnostic measures by comparing their ROC areas, or to
establish for a single measure the tradeoff between data
length and misidentifications (misses and false positives) by
examining ROC area as a function of record length.
A minimum record length can then be specified to achieve acceptable
classification accuracy.

As pointed out above, ROC analysis relies on no implicit assumptions
about the statistical nature of the data set
\cite{vantrees68,swets88}, so that it is generally more suitable \cite{herz2}
for analyzing non-Gaussian time series than are measures
of statistical significance such as {\em p}-value, $h$, and $d$.
Another important feature of ROC curves is that they
are insensitive to the units employed (e.g.,
spectral magnitude, magnitude squared, or log magnitude); ROC curves
for a measure $M$ are identical to those for any monotonic
transformation thereof such as $M^x$ or $\log(M)$.
In contrast the values of $d', h$, and $d$ are generally modified by
such transformations, as will be demonstrated in
Sec.~\ref{sssec:compddm}.

\section{Discriminating Heart-Failure Patients from Normal Subjects}
\label{sec:discrim}

We now proceed to examine the relative merits of various HRV measures
for discriminating congestive-heart-failure (CHF) patients from
normal subjects.
Specifically we contrast and compare the performance of the 16
measures set forth in Sec.~\ref{sec:methandmeas}: VLF, LF, HF,
LF/HF, pNN50, SDANN, SDNN ($\sigma_{\rm int}$), $A(T)$,
$\sigma_{\rm wav}(m)$, $S_{\tau}(f)$, DFA($m$), $\alpha_D$,
$\alpha_W$, $\alpha_S$, $\alpha_A$, and $\alpha_R$.

After discussing the selection of an appropriate scale $m$,
we use predictive
value plots and ROC
curves to select a particular subset of HRV markers that appears to
be promising for discerning the presence of
heart failure in a patient population.

\subsection{Database}
\label{ssec:database}

The RR recordings analyzed in this section were drawn from
the Beth-Israel Hospital (Boston, MA) Heart-Failure
Database which includes 12 records
from normal subjects (age 29--64 years, mean 44 years)
and 12 records from severe congestive-heart-failure patients (age
22--71 years, mean 56 years).
The recordings were made with a Holter monitor digitized at a fixed
value of 250 samples/sec.
Also included in this database are 3 RR records for CHF patients who
also suffered from atrial fibrillation (AF); these records are
analyzed as a separate class.
All records contain both diurnal and nocturnal segments.
The data were originally provided to us in 1992 by D.~Rigney and
A.~L.~Goldberger.

A detailed characterization of each of the records is presented in
Table~1 of Ref.~\cite{turcott96}; some statistical details are
provided in Table~A1.
Of the 27 recordings, the shortest contained $L_{\rm max}=75821$ RR
intervals; the remaining 26 recordings were truncated to this length
before calculating the 16 HRV measures.

\subsection{Selecting a Scale}

A value for the scale $m$ that suitably discriminates heart-failure
patients from normal subjects can be inferred from our recent wavelet
studies of the CHF and normal records from the same database as
discussed in Sec.~\ref{ssec:database} \cite{thurner98,herz2}.
With the help of the wavelet-transform standard deviation
$\sigma_{{\rm wav}}(m)$ discussed in detail in
Sec.~\ref{sssec:wtsddef}, we discovered a critical scale window
near $m=32$ interbeat intervals over which the normal subjects
exhibited greater fluctuations than those afflicted with heart
failure.
For these particular long data sets, we found that it was possible to
perfectly discriminate between the two groups
\cite{thurner98,herz2,teich98}.

The  results are displayed in Fig.~4, where
$\sigma_{\rm wav}(m)$ is plotted vs. wavelet scale $m$
for the 12 normal subjects ($+$), the 12 CHF patients $\bar{\rm{s}}$
atrial fibrillation ($\times$), and the 3 CHF patients $\bar{\rm{c}}$
atrial fibrillation ($\triangle$), using Haar-wavelet analysis.
The AF patients ($\triangle$) typically fell near the high end of the
non-AF patients ($\times$), indicating greater RR fluctuations,
particularly at small scales.
This results from the presence of non-sinus beats.
Nevertheless it is evident from Fig.~4 that the wavelet measure
$\sigma_{{\rm wav}}$ serves to completely separate the normal
subjects from the heart-failure patients (both $\bar{\rm {s}}$ and
$\bar{\rm {c}}$ AF) at scales of 16 and 32 heartbeat intervals, as
reported in Ref. \cite{thurner98}.
One can do no better.
This conclusion persists for a broad range of analyzing wavelets,
from Daubechies 2-tap (Haar) to Daubechies 20-tap \cite{thurner98}.

The importance of this scale window has been recently confirmed in an
Israeli-Danish
study of diabetic patients who had not yet developed clinical signs
of cardiovascular disease \cite{ashkenazy98}.
The reduction in the value of the wavelet-transform standard
deviation $\sigma_{{\rm wav}}(32)$ that leads to the scale window
occurs not only for CHF ($\bar{\rm{s}}$ and $\bar{\rm{c}}$ AF) and
diabetic patients, but also for heart-transplant patients
\cite{teich98,ashkenazy98}, and also in records preceding sudden cardiac
death \cite{thurner98,teich98}.
The depression of $\sigma_{{\rm wav}}(32)$ at these
scales is likely associated with the impairment of
autonomic nervous
system function. Baroreflex modulations of the sympathetic or
parasympathetic tone typically lie in the range 0.04--0.09 cycles/sec
(11--25 sec), which corresponds to the scale where
$\sigma_{{\rm wav}}(m)$ is reduced.

These studies, in conjunction
with our earlier investigations which revealed a similar critical
scale window in the {\em counting} statistics of the
heartbeat \cite{turcott96,teich96c}
(as opposed to the time-{\em interval} statistics under discussion),
lead to the recognition that scales in the vicinity of $m=32$ enjoy a
special status. Those measures that depend on a particular scale are
therefore evaluated at $m=32$ and $f=1/32$ in the expectation that
these values maximize discriminability in the more usual situation when
the two classes of data cannot be fully separated.

\subsection{Individual Value Plots}
\label{ssec:ivplots}

Having devised a suitable scale value $m$ we now proceed to evaluate the 16
measures for all 27 normal and CHF data sets, each comprising
$75821$ RR intervals. The results are presented
in Fig.~5 where each of the 16
panels represents a different measure. For each measure the individual
values for normal subjects ($+$), CHF patients $\bar{\rm {s}}$ AF
($\times$), and CHF
patients $\bar{\rm {c}}$ AF ($\triangle$) comprise the left three columns,
respectively. Values in the right four columns correspond
to other cardiovascular pathologies and will be discussed
in Sec.~\ref{sec:markers}.

To illustrate how particular measures succeed (or fail to succeed) in
distinguishing between CHF patients and normal subjects, we focus in
detail on two measures: VLF power and pNN50.
For this particular collection of patients and record lengths, the
normal subjects all exhibit larger values of VLF power than do the
CHF patients; indeed a horizontal line drawn at VLF$=0.000600$
completely separates the two classes.
On the other hand for pNN50, though the normals still have larger
values on average, there is a region of overlap of CHF patients and
normal subjects near 0.05, indicating that the two classes of
patients cannot be entirely separated using this measure.
Thus for the full data set, comprising
$75821$ RR intervals, VLF succeeds in completely distinguishing
CHF patients and normal subjects whereas pNN50 does not.

Examining all 16 panels, we find that six measures manage to
completely separate the normal subjects (first column) from the
heart-failure patients (second and third columns) while the
remaining 10 fail to do so.
The six successful measures are highlighted by boldface font in
Fig.~5: {\boldmath {\bf VLF}, {\bf LF}, $A(10)$,
$\sigma_{\rm wav}(32)$, $S_{\tau}(1/32)$, and {\bf DFA(32)}.}

\subsection{Predictive Value Plots}
\label{ssec:pvplots}

How can the ability of a measure to separate two classes
of subjects be quantified? Returning to the VLF panel in Fig.~5
we place a threshold level $\theta$ at an arbitrary position
on the ordinate, and consider only the leftmost
two columns: normal subjects and heart-failure
patients who do not suffer from atrial fibrillation.
We then classify all subjects for whom the VLF values are
$<\theta$ as CHF patients (positive) and those
for whom the VLF
values $> \theta$ as normal (negative). (Measures
that yield
smaller results for normal patients, on average, obey a reversed
decision criterion.)

If a subject labeled as a CHF patient is indeed so afflicted, then
this situation is referred to as a true positive ($P_T$); a normal
subject erroneously labeled as a CHF patient is referred to as a
false positive ($P_F$).
We define negative outcomes that are true ($N_T$) and false ($N_F$)
in an analogous manner.
As pointed out in Sec.~\ref{sssec:pvdef}, the positive
predictive value $V_{P} = P_T/(P_T+P_F)$ and negative predictive
value $V_{N} = N_T/(N_T+N_F)$ represent the proportion of positives
and negatives, respectively, that are correctly identified.
This determination is carried out for many values of the threshold
$\theta$.

Figure~6 shows the positive (solid curves) and negative (dotted
curves) predictive values for all 16 measures, plotted against the
threshold $\theta$, each in its own panel.
These curves are constructed using the 12 normal and 12 heart-failure
($\bar{\rm{s}}$ AF) records that comprise the CHF database discussed
in Sec.~\ref{ssec:database}.
For the VLF measure, both predictive
values are simulataneously unity in the immediate vicinity of
$\theta = 0.000600$. This occurs because $P_F$ and $N_F$ are
both zero at this particular
value of $\theta$ and reconfirms that the two classes of data
separate perfectly in the VLF panel of Fig.~5
at this threshold.

For threshold values outside the range
$0.000544 < \theta < 0.000603$, some of the patients will be
incorrectly identified by the VLF measure. If we set
$\theta = 0.000100$, for
example, six of the twelve CHF patients will be incorrectly
identified as normal subjects, which is confirmed by examining
the VLF panel in Fig.~5. This yields
$V_{N} = N_T/(N_T+N_F) = 12/(12 + 6) \doteq 0.67 < 1 $,
which is the magnitude of the negative predictive value
(dotted curve) in the VLF
panel in Fig.~6. At this value of
the threshold ($\theta = 0.000100$) the positive predictive value remains unity
because $P_F$ remains zero.

The pNN50 panel in Fig.~5, in contrast, reveals a range of
overlap in the individual values of the normal subjects
and CHF patients. Consequently as $\theta$ increases
into the overlap region, $V_{P}$ decreases below unity and this
happens before $V_{N}$
attains unity value. Thus there is no threshold value, or range
of threshold values,
for which the positive and negative predictive values
in Fig.~6 are both unity.
The best threshold for this measure lies
in the range $0.026 < \theta < 0.050$, with the choice depending
on the relative benefit of being able to accurately predict the
presence or absence of CHF in a patient.

There are six measures in Fig.~6 (indicated in boldface font) for which
the positive and negative predictive values are both
unity over the same range of threshold values. These
measures are, of course, the same six measures for
which the normal subjects and heart-failure patients fall into
disjoint sets in Fig.~5.

\subsection{ROC Curves}
\label{ssec:roccurves}

Two other important clinically relevant quantities that depend on the
threshold $\theta$ are the
sensitivity, the proportion of heart-failure patients that are properly
identified [$P_T/(P_T+N_F)$]; and the specificity, the proportion of
normal subjects that are properly identified [$N_T/(N_T+P_F)$].
As pointed out in subsections~\ref{sssec:pvdef}
and~\ref{sssec:rocdef}, sensitivity and specificity relate to patient
status (pathologic and normal, respectively) whereas the positive and
negative predictive values relate to identification status (positive
and negative, respectively).
Sensitivity and specificity are both monotonic functions of the
threshold, but this is not generally true for the predictive values.
The monotonicity property is salutary in that it
facilitates the use of a
parametric plot
which permits these quantities to be represented in compact form.
A plot of sensitivity versus $1 -$specificity,
traced out using various values of
the threshold $\theta$, forms the receiver-operating-characteristic
(ROC) curve (see Sec.~\ref{sssec:rocdef}).

ROC curves are presented in Fig.~7 for all 16 measures, again using
the same 12 normal and 12 heart-failure ($\bar{\rm{s}}$ AF) records
that comprise the CHF database discussed in Sec.~\ref{ssec:database}.
Because of the complete separation between the two classes of
patients (leftmost two columns of the VLF panel in Fig.~5) near
$\theta = 0.000600$, the VLF ROC curve in Fig.~7 simultaneously
achieves unity (100\%) sensitivity and unity (100\%) specificity (the
point at upper left corner of the ROC curve).
For the pNN50 statistic, in contrast, the overlap evident in Fig.~5
prevents this, so that the upper left corner of the pNN50 ROC curve
in Fig.~7 instead reveals smaller simultaneous values of sensitivity
and specificity.

Six measures in Fig.~7 simultaneously exhibit unity sensitivity and
specificity; these are indicated by boldface font and have ROC curves
that are perfectly square.
They are clearly the same measures for which the normal subjects and
heart-failure patients fall into disjoint sets in Fig.~5, and for
which simultaneous positive and negative predictive values of unity
are observed in Fig.~6.

\subsubsection{Comparison with Detection-Distance Measures}
\label{sssec:compddm}

For didactic purposes we compare the ROC results presented
immediately above with those obtained using detection-distance
analysis.
As we indicated in Sec.~\ref{sssec:statsig}, care must be exercised
when using these techniques for anything other than
Gaussian-distributed quantities.
The calculations were carried out using the same 12 normal-subject
and 12 CHF-patient records, each comprising $L_{\rm max}=75821$
intervals.
In Table~1 we provide the detection distances $h$ and $d$, in order
of descending value of $h$, for all 16 measures.
Large values are best since they indicate that the two distributions
are well separated.

Five of the six measures that entirely separate the CHF patients and
normal subjects using ROC analysis fall in the top five positions in
Table~1.
The sixth measure, LF, falls in ninth position.
This confirms that detection-distance analysis applied to these long
recordings provides results that qualitatively agree
with those obtained using ROC analysis. However, detection-distance
analysis does not provide any indication of
how many (or indeed whether any) of the
measures at the top of the list completely separate
the two classes of patients, nor does it provide
estimates of sensitivity and specificity. Moreover, the rankings
according to $d$ differ from those according to $h$.

Finally, the detection-distance for a particular
measure, as well as the relative ranking of that measure, depends on
what appear to be insignificant details about the specific form
in which the measure is cast. For example the $h$ values
for $\sigma_{\rm wav}(32)$ and its square
$\sigma^{2}_{\rm wav}(32)$ are substantially different, and so are
their rankings in Table~1. As discussed in Sec.~\ref{sssec:rocdef}, ROC
analysis is invariant to monotonic transformations of the measure
and therefore does not suffer from this disadvantage.

\subsubsection{ROC-Area Curves}

Perfectly square ROC curves,
associated with the group of six boldface-labeled
measures in Figs.~5--8,
exhibit unity area. These ROCs represent 100\%
sensitivity for all values of specificity,
indicating that every patient is properly assigned
to the appropriate status: heart-failure or normal.
Though the perfect separation achieved by these six measures
endorses them as useful diagnostic statistics, the
results of most studies are seldom so clear-cut.
ROC area will surely decrease as increasing numbers
of out-of-sample records are added to the database, since
increased population size means increased variability
\cite{thurner98}.

ROC area also decreases with diminishing data length; shorter
records yield less information about patient
condition and these patients are therefore more likely to be misclassified
\cite{herz2,teich98}. The ROC curves in Fig.~7 have been
constructed from
Holter-monitor records that contain many hours
of data ($75821$ RR intervals). It would be useful in a clinical setting
to be able to draw inferences from HRV measures recorded
over shorter times, say minutes rather than hours.
It is therefore important to
examine the performance of the 16 HRV
measures as data length is decreased. As indicated in Sec.~\ref{sssec:rocdef},
ROC-analysis provides an ideal method for carrying out this task
\cite{herz2,teich98}.

In Fig.~8 we present ROC-area curves as a function of the number of
RR intervals (data length) $L$ analyzed ($64 \leq L \leq 75821$).
The ROC areas for the full-length records ($L_{\rm max}=75821$),
which correspond to the areas under the ROC curves presented in
Fig.~7, are the rightmost points in the ROC-area curves shown in
Fig.~8.
Results for shorter records were obtained by dividing the 12 normal
and 12 heart-failure ($\bar{\rm{s}}$ AF) records that comprise the
CHF database (Sec.~\ref{ssec:database}) into smaller segments of
length $L$.
The area under the ROC curve for that data length $L$ was computed
for the first such segment for all 16 measures, and then for the
second segment, and so on, for all segments of length $L$ (remainders
of length $<L$ of the original files were not used for the ROC-area
calculations).
>From the $L_{\rm max} / L$ values of the ROC area, the mean and
standard deviation were computed and plotted in Fig.~8.
The lengths $L$ examined ranged from $L=2^6=64$ to $L=2^{16}=65536$
RR intervals, in powers of two, in addition to the entire record of
$L_{\rm max}=75821$ intervals.

To illustrate the information provided by
these curves, we direct our attention to
the VLF and pNN50 panels in Fig.~8.
For the full-length records the
right-most point in the VLF panel reveals unity area while that for
pNN50 lies somewhat lower, as expected from the
corresponding ROC
curves in Fig.~7. VLF clearly outperforms pNN50. As the data length
analyzed decreases, so too do the ROC areas for
both measures while their variances increase, also as expected. However,
when the data
length dips to $256$ or fewer RR intervals, the performance
of the two measures reverses so that pNN50 outperforms VLF.
There is an important point to be drawn from this example. Not
only does the performance of a measure depend
on data length,
but so too does the relative performance of different measures.

\subsection{Comparing the Measures for Various Data Lengths}
\label{ssec:compmeas}

Based on their overall ability to distinguish between CHF patients
and normal subjects over a range of data lengths, the sixteen
measures shown in Fig.~8 divide roughly into three classes.
The six measures that fall in the first class, comprising VLF, LF,
$A(10)$, $\sigma_{\rm wav}(32)$, $S_{\tau}(1/32)$, and DFA(32),
succeed in completely separating the two classes of patients for data
lengths down to $L=2^{15} = 32768$ RR intervals.
These six measures share a dependence on a single scale, or small
range of scales, near 32 heartbeat intervals.
For this collection of data sets, this scale appears to yield the
best performance.
Members of this class outperform the other ten measures at nearly all
data lengths.
Apparently the scale value itself is far more important than the
measure used to evaluate it.

The second class, consisting of HF, the ratio LF$/$HF, pNN50, and
$\sigma_{\rm int}$, fail to achieve complete separation for any data
size examined. Nevertheless the members of this class are
not devoid of value in separating
CHF patients from normal subjects. Interestingly, all but LF$/$HF
provide better results than
$A(10)$, a member of the first class,
for the shortest data lengths. Results for these four measures
varied relatively little with data size, thus exhibiting a form of
robustness.

Members of the third class, consisting of SDANN and the
five scale-independent
measures $\alpha_{[\cdot]}$, exhibit poor performance at all
data lengths. These six
measures require long sequences of RR intervals to make available
the long-term fluctuations required for accurate estimation of the
fractal exponent. Data lengths $L<5000$ RR intervals
lead to large variance and (negative) bias, and
are not likely to be meaningful. As an example of
the kind of peculiarity
that can emerge when attempting to apply
scale-independent measures to short records, the $\alpha_A$ ROC area
decreases below 0.5 when the data size falls below 2048 intervals
(reversing the sense of the comparison only for these data sizes
increases the ROC area, though not above 0.7; however this
clearly violates
the spirit of the method).
SDANN requires several 5-minute segments to
accurately determine the standard deviation.

\subsubsection[Scale-Independent $vs$ Scale-Dependent
Measures]{Scale-Independent {\boldmath $vs$} Scale-Dependent
Measures}
\label{sssec:sivssd}

As indicated in the previous subsection, all five scale-independent
measures ($\alpha_D$, $\alpha_W$, $\alpha_S$, $\alpha_A$,
and $\alpha_R$)
perform poorly at all data lengths.
These fractal-exponent estimators return widely differing
results as is plainly evident in Fig.~5. This suggests
that there is little merit in the concept
of a single exponent for
characterizing the human heartbeat sequence, no less a ``universal'' one
as some have proposed \cite{peng95,peng93,amaral98}.

A variation on this theme is
the possibility that pairs of fractal exponents
can provide a useful HRV measure.
At small scales $m$, Fig.~4 reveals that
heart-failure patients exhibit smaller values of the
wavelet-transform standard-deviation slope than do normal subjects.
Following Peng {\em et al.}
\cite{peng95}, who constructed a measure based on
differences of DFA scaling exponents in different scaling regions
in an attempt
to discriminate CHF patients from normal subjects,
Thurner {\em et al.} \cite{thurner98}
constructed a measure based on differences in the
wavelet-transform standard-deviation slope at different
scales.
However
the outcome was found to be unsatisfactory when compared with other
available measures; we concluded the same about
the results obtained by Peng {\em et al.} \cite{peng95}.
Using ROC analysis, as described in Sec.~\ref{sssec:rocdef}, we
determined that the ROC area for the measure described by Thurner {\em et al.}
\cite{thurner98} was sufficiently small
(0.917 for $m=4, 16,$ and
$256$) that we abandoned this construct.

Four of the techniques we have discussed in this Chapter (spectral,
wavelet, detrended fluctuation analysis, and Allan factor)
yield both scale-independent and scale-dependent
measures and therefore afford us the opportunity of directly
comparing these two classes of measures in individual calculations:
$\alpha_W \leftrightarrow \sigma_{\rm wav}(32)$;
$\alpha_S \leftrightarrow S_{\tau}(1/32)$;
$\alpha_D \leftrightarrow$ DFA(32);
$\alpha_A \leftrightarrow A(10)$.
In each of these cases the fixed-scale measure is found to
greatly outperform the fractal-exponent measure for all data sizes
examined, as we previously illustrated for the pairs
$\alpha_W \leftrightarrow \sigma_{\rm wav}(32)$ and
$\alpha_S \leftrightarrow S_{\tau}(1/32)$ \cite{herz2}.
These results were recently confirmed in a follow-up Israeli--Danish
study \cite{ashkenazy99}.
Moreover, in contrast with the substantial variability returned in
fractal-exponent estimates, results for the
different scale-dependent measures at $m=32$ intervals
bear reasonable similarity to each other.

Nunes Amaral {\em et al.} \cite{amaral98} recently concluded exactly
the opposite,
namely that scaling exponents provide superior performance
to scale-dependent measures.
This may be because they relied exclusively on the
distribution-dependent measures $\eta \equiv h^2$ and $d^2$
(see subsections~\ref{sssec:statsig} and \ref{sssec:compddm})
rather than on distribution-independent ROC analysis.
These same authors \cite{amaral98} also purport to glean information
from higher moments of the wavelet coefficients, but the reliability
of such information is questionable because estimator variance
increases with moment
order \cite{thurner97}.

\subsubsection{Computation Times of the Various Measures}
\label{sssec:comptimes}

The computation times for the 16 measures considered in this
Chapter are provided in Table~2. All measures
were run ten times and averaged except for the two
DFA measures which, because of their long execution time,
were run only once. These long execution times are associated with
the suggested method for
computing DFA \cite{buldyrev95}, which is an $N^2$
process. DFA computation times
therefore
increase as the square of the number of intervals whereas
all 14 other methods, in contrast, are either of order $N$ or
$N \log(N)$.

Based on computation time, we can rank order the six scale-dependent
measures that fall into the first class, from fastest to slowest:
$\sigma_{\rm wav}(32), S_{\tau}(1/32), A(10),$ LF, VLF, and DFA(32).
Because of its computational simplicity the wavelet-transform
standard deviation $\sigma_{\rm wav}(32)$ computes more rapidly than
any of the other measures.
It is 3 times faster than that of its nearest competitor
$S_{\tau}(1/32)$, 16.5 times faster than LF, and $32500$ times faster
than DFA(32).

\subsubsection{Comparing the Most Effective Measures}
\label{sssec:mosteff}

In subsections~\ref{ssec:ivplots}--\ref{ssec:roccurves}, we
established that six measures succeeded in completely separating the
normal subjects from the CHF patients in our database for data
lengths $L \geq 2^{15} = 32768$ RR intervals: VLF, LF, $A(10)$,
$\sigma_{\rm wav}(32)$, $S_{\tau}(1/32)$, and DFA(32).
We now demonstrate from a fundamental point of view that these
measures can all be viewed as transformations of the interval-based
power spectral density and that all are therefore closely related.

Consider first the relationships of VLF and LF to $S_{\tau}(1/32)$.
At a frequency $f = 1/32 \approx 0.031$ cycles/interval, $S_{\tau}(f)$
separates CHF patients from normal subjects; however, a range of
values of $f$ provide varying degrees of separability.
For the data sets we analyzed, separation extends from $f=0.02$ to
$f=0.07$ cycles/interval \cite{heneghan98}.
Recall that VLF and LF are simply integrals of $S_{\tau}(f)$ over
particular ranges of $f$: from $f=0.003$ to $f=0.04$ cycles/interval
for VLF, and from $f=0.04$ to $f=0.15$ cycles/interval for LF.
Since these ranges overlap with those for which the power spectral
density itself provides separability, it is not surprising
that these spectral integrals also exhibit this property.
This illustrates the close relationship among VLF, LF, and
$S_{\tau}(1/32)$.

We next turn to the relation between $\sigma_{\rm wav}(32)$ and
$S_{\tau}(1/32)$.
In Sec.~\ref{sssec:wtandspec} the relationship between these
two measures was established as
\begin{equation}
\sigma_{\rm wav}^2(m)=m \int^\infty_{-\infty}S_{\tau}(f)H(mf)df,
\end{equation}
revealing that $\sigma^{2}_{\rm wav}(m)$ is simply a
bandpass-filtered version of the interval-based power spectral
density for stationary signals.
For the Haar wavelet, the bandpass filter $H(mf)$ has a center
frequency near $1/m$, and a bandwidth about that center frequency
also near $1/m$ \cite{heneghan98}.
Accordingly, $\sigma^{2}_{\rm wav}(32)$ is equivalent to a weighted
integral of $S_{\tau}(f)$ centered about $f = 1/32 \approx 0.031$,
with an approximate range of integration of $0.031 \pm 0.016$
(0.016 to 0.047).
Thus the separability of $\sigma^{2}_{\rm wav}(32)$ is expected to
resemble that of $S_{\tau}(1/32)$, and therefore that of VLF and LF
as well.

Now consider the count-based Allan factor evaluated at a counting
time of ten seconds, $A(10)$.
Since the Allan factor is directly related to the variance of the
Haar wavelet transform of the {\it counts} \cite{teich96a}, $A(10)$
is proportional to the wavelet transform of the counting process
evaluated over a wavelet duration of $T=20$ sec.
The count-based and interval-based measures are approximately related
via Eq.~(\ref{eq:freqconv}) so that $A(10)$ should be compared with
$\sigma_{\rm wav}(20/{\rm E}[\tau])\approx \sigma_{\rm wav}(25)$
whose value, in turn, is determined by $S_{\tau}(1/25 = 0.04)$ and
spectral components at nearby frequencies.
Since this range of $S_{\tau}(f)$ is known to exhibit separability,
so too should $A(10)$.

The relationship between detrended fluctuation analysis and the other
measures proves more complex than those delineated above.
Nevertheless, DFA(32) can also be broadly interpreted in terms of an
underlying interval-based power spectral density.
To forge this link, we revisit the calculation of $F(m)$ (see
Sec.~\ref{sssec:dfadef}).
The mean is first subtracted from the sequence of RR intervals
$\{\tau_i\}$ resulting in a new sequence that has zero mean.
This new random process has a power spectral density everywhere equal
to $S_{\tau}(f)$, except at $f=0$ where it assumes a value of zero.
The summation of this zero-mean sequence generates the series $y(k)$,
as shown in Eq.~(\ref{eq:sumser}).
Since summation is represented by a transfer function $1/j2\pi f$ for
small frequencies $f$, with $j=\sqrt{-1}$, the power spectral density
of $y(k)$ is approximately given by
\begin{equation}
S_y(f) \approx \left\{ \begin{array}{ll}
0 & \mbox{$f = 0$} \\
\frac{\displaystyle{S_{\tau}(f)}}{\displaystyle{4\pi^2f^2}} &
\mbox{$f \neq 0$.}
\end{array} \right.
\end{equation}

Next, the sequence $y(k)$ is divided into segments of length $m$.
For the first segment, $k=1$ to $m$, the best fitting linear trend
$y_m(k) = \beta + \gamma k$ is obtained; the residuals $y(k) -
y_m(k)$ therefore have the minimum variance over all choices of
$\beta$ and $\gamma$.
Over the duration of the entire sequence $y(k)$, the local trend
function $y_m(k)$ will consist of a series of such linear segments,
and is therefore piecewise linear.
Since $y_m(k)$ changes behavior over a time scale of $m$ intervals,
its power will be concentrated at frequencies below $1/m$.
Since the residuals $y(k) - y_m(k)$ have minimum variance, the
spectrum of $y_m(k)$ will resemble that of $y(k)$ as much as
possible, and in particular at frequencies below $1/m$ where most of
its power is concentrated.
Thus
\begin{equation}
S_{ym}(f) \approx \left\{ \begin{array}{ll}
S_y(f) & \mbox{$|f| < 1/m$} \\
0 & \mbox{$|f| > 1/m$.}
\end{array} \right.
\end{equation}

Equation~(\ref{eq:dfadef}) shows the mean-square fluctuation $F^2(m)$
to be an estimate of the variance of the residuals $y(k) - y_m(k)$.
By Parseval's theorem, this variance equals the integral of the
spectrum of the residuals over frequency.
Therefore
\begin{equation}
F^2(m) \approx 2 \int_{0^+}^{\infty}[S_y(f) - S_{ym}(f)]df
= (2\pi^2)^{-1} \int_{1/m}^{\infty} S_{\tau}(f) f^{-2}df
\label{eq:dfaaprx}
\end{equation}
so that the mean-square fluctuation $F^2(m)$ can be represented as a
weighted integral of the RR-interval spectrum over the appropriate
frequency range.

We conclude that all six measures that best distinguish CHF patients
from normal subjects are related in terms of transformations of the
interval-based power spectral density.

\subsubsection{Selecting the Best Measures}

In the previous section we showed that the six most effective
measures are interrelated from a fundamental point of view.
This conclusion is confirmed by the overall similarity of the
ROC-area curves for these measures, as can be seen in Fig.~8.

We now proceed to compare these ROC areas more carefully
in Fig.~9. This figure presents
$1-$ROC area plotted against the data length analyzed
$L$ on a doubly logarithmic plot. This is in contrast to the usual
presentation (such as Fig.~8) in which ROC area is plotted against $L$
on a semilogarithmic plot. The unconventional form of the plot
in Fig.~9 is designed to allow small
deviations from unity area to be highlighted.
Clearly, small
values of the quantity $1-$ROC area are desirable. Of course, as the
data length $L$ diminishes, ROC area decreases (and therefore
$1-$ROC area increases) for all six measures.

The trends exhibited by all six measures displayed in Fig.~9
(VLF, $A(10)$, DFA(32),
$\sigma_{\rm wav}(32)$, $S_{\tau}(1/32)$, and LF) are indeed
broadly similar, as we observed in Fig.~8.
For data sets that are sufficiently long, the six measures provide
identical diagnostic capabilities (unity ROC area).
At shorter data lengths, however, VLF, $A(10)$, and DFA(32) exhibit
more degradation than do the other three measures.
In accordance with the analysis provided in
Sec.~\ref{sssec:mosteff}, VLF deviates the most at small values
of $L$ because it incorporates information from values of
$S_{\tau}(f)$ well below the frequency range of separability.
Similarly DFA(32) favors low-frequency contributions by virtue of the
$1/f^2$ weighting function in Eq.~(\ref{eq:dfaaprx}).
Finally, the performance of $A(10)$ is suppressed because of the
confounding effect of values of $S_{\tau}(f)$ for $f>0.07$, as well
as from possible errors deriving from our assumptions regarding the
equivalence of the count-based and interval-based wavelet-transform
variances.

The remaining three measures therefore emerge as superior:
LF, $\sigma_{\rm wav}(32)$, and $S_{\tau}(1/32)$.
Judged on the basis of both performance (Fig.~9) {\em and}
computation time (Table~2), we conclude that two of these are best:
{\boldmath $\sigma_{\rm wav}(32)$ and $S_{\tau}(1/32)$}.

It is desirable to confirm these conclusions for other
records, comprising out-of-sample data sets from CHF patients
and normal subjects. It will also be
of interest to examine the correlation of ROC area with
severity of cardiac dysfunction as judged, for example,
by the NYHA clinical scale. It is also important to conduct
comparisons with other CHF studies that make use of
HRV measures \cite{malik96}.

A sufficiently large database of normal and CHF
records would make it
possible to examine the detailed distinctions
between the two classes
of patients over a range of scales surrounding $m=32$. A superior
measure
which draws on these distinctions in an
optimal way could then be designed; for example the weighting function in
an integral over $S_{\tau}(f)$ could be customized.
Other measures that simultaneously
incorporate properties inherent in collections of the
individual
measures considered here could also be developed.

\section{Markers for Other Cardiac Pathologies}
\label{sec:markers}

Returning to the individual-value plots in Fig.~5, we briefly examine
the behavior of the 16 measures for
several other cardiovascular disorders for which these measures may
hold promise.
Among these are three CHF patients who also suffer from
atrial
fibrillation (CHF $\bar{\rm{c}}$ AF, third column, $\triangle$),
one heart-transplant patient (TRANSPLANT, fourth column,
\hbox{$\times$ \hspace{-1.45em} $+$}),
one ventricular-tachycardia patient (VT, fifth column, $\diamondsuit$),
one sudden-cardiac-death patient (SCD, sixth column,
\hbox{$\sqcup$ \hspace{-1.35em} $\sqcap$}), and
two sleep-apnea patients (APNEA, last column, $+$).
A summary of the more salient statistics of the original
RR recordings for these patients is presented in Table~A1.
Prior to analysis, the RR records were truncated at
$L_{\rm max}=75821$ RR intervals to match the lengths of the
CHF-patient and normal-subject records.
The two sleep-apnea data records were drawn from the
MIT-BIH Polysomnographic Database (Harvard-MIT Division of Health
Sciences and Technology) and comprised $16874$ and $15751$ RR
intervals respectively. Because of the limited
duration of the sleep period, records containing larger numbers of RR
intervals are not available for this disorder.

Too few patient
recordings are available to us to reasonably evaluate
the potential of these HRV measures for the diagnosis of
these forms of cardiovascular disease, but the results presented
here might suggest which measures would prove useful given
large numbers of records.

We first note that for most of the 16 measures,
the values for the three CHF patients with atrial fibrillation
($\triangle$)
tend to fall within the range set by the other
twelve CHF patients without atrial fibrillation. In particular
the same six measures that completely separate normal subjects
from CHF patients $\bar{\rm s}$ AF continue to do so when CHF
patients $\bar{\rm c}$ AF are included.
The inclusion of the AF patients reduces the
separability only for three of the 16 measures: pNN50, HF, and
$\alpha_W$.

Results for the heart transplant patient
(\hbox{$\times$ \hspace{-1.45em} $+$})
appear towards the pathological end of the CHF range for many
measures, and for some measures the transplant value extends beyond the
range of all of the CHF patients. Conversely, the sleep apnea
values ($+$) typically fall at the end of, or beyond, the range
spanned by the normal subjects.
Indeed, three thresholds can be chosen for
$S_{\tau}(1/32)$ (i.e., 0.000008, 0.0006, and 0.00436)
which completely separate four classes of patients:
sleep apnea [largest values of $S_{\tau}(1/32)$], normal subjects
(next largest), CHF with or without atrial fibrillation (next
largest), and heart transplant (smallest).
While such separation will no doubt disappear as
large numbers of additional
patient records are included, we nevertheless expect the largest
variability
(measured at 32 intervals as well as at other scales)
to be associated with
sleep-apnea patients since their fluctuating breathing patterns
induce greater fluctuations in the heart rate. Conversely,
heart-transplant
patients (especially in the absence of reinnervation) are expected to
exhibit the lowest values of variability since the autonomic
nervous system, which contributes to the modulation of heart rate
at these time
scales, is subfunctional.

Results for the ventricular tachycardia patient ($\diamondsuit$) lie within the
normal range for most measures, though the value for LF$/$HF is
well beyond the pathological end of the range for CHF patients.
Interestingly, four measures [VLF, HF, SDANN, and $A(10)$] yield the
largest values for the ventricular tachycardia patient of the 32
data sets examined.
Perhaps $A(10)$, which exhibits the largest separation between the
value for this patient and those for the other 31 patients, will prove
useful in detecting ventricular tachycardia.

Finally we consider results for the sudden-cardiac-death patient
(\hbox{$\sqcup$ \hspace{-1.35em} $\sqcap$}),
which lie between the ranges of normal subjects and CHF patients for all but
one (LF) of the six measures which separate these two patient
classes. The pNN50 and HF results, on the other hand,
reveal values
beyond the normal set, thereby indicating the presence of greater
variability on a short time scale than for other patients.
The value of SDANN for the SCD patient is greater than that for any
of the other 31 patients so it is possible that SDANN could
prove useful in predicting sudden cardiac death.

\section{Does Deterministic Chaos Play a Role in Heart Rate Variability?}

The question of whether normal HRV arises from a low-dimensional
attractor associated with a deterministic dynamical system or has
stochastic underpinnings has continued to entice researchers
\cite{bassingthwaighte94a}.
Previous studies devoted to this question have come to conflicting
conclusions.
The notion that the normal human-heartbeat sequence is chaotic
appears to have been first set forth by Goldberger and West
\cite{goldberger87}.
Papers in support of this hypothesis \cite{poon97}, as well as in opposition to
it \cite{kanters94,roach98}, have appeared recently.
An important factor that pertains to this issue is the recognition
that correlation, rather than nonlinear dynamics, can lie behind
these seemingly chaotic recordings \cite{turcott96}.

We address this question in this section and conclude, for
both normal subjects and patients with several forms
of cardiac dysfunction, that the sequence of heartbeat
intervals does not exhibit chaos.

\subsection{Methods}

\subsubsection{Phase-Space Reconstruction}

One way of approaching the question of chaos in heartbeat recordings
is in terms of a phase-space reconstruction and the use of an
algorithm to estimate one or more of its fractal dimensions $D_q$
\cite{bassingthwaighte94b,schepers92,theiler90}.
The box-counting algorithm \cite{liebovitch89b} provides a method for
estimating the capacity (or box-counting) dimension $D_0$ of the
attractor \cite{bassingthwaighte94b,turcott96,theiler90,ding93}.

The phase-space reconstruction approach operates on the basis that
the topological properties of the attractor for a dynamical system
can be determined from the time series of a single observable
\cite{grassberger83,liebovitch89a,wolf85}.
A $p$-dimensional vector
$\stackrel{\textstyle \rightarrow}{T} \equiv
\{\tau_i, \tau_{i + l}, ... \tau_{i + (p-1)l} \}$
is formed from the sequence of RR intervals $\{\tau_i\}$.
The parameter $p$ is the embedding dimension, and $l$ is the lag,
usually taken to be the location of the first zero crossing of the
autocorrelation function of the time series under study.
As the time index $i$ progresses, the vector
$\stackrel{\textstyle \rightarrow}{T}$
traces out a trajectory in the $p$-dimensional embedding space.

The box-counting algorithm estimates the capacity dimension of this
trajectory.
Specifically, the negative slope of the logarithm of the number of
full boxes versus the logarithm of the width of the boxes, over a range
of $4/[\max(\tau) - \min(\tau)]$ to $32/[\max(\tau) - \min(\tau)]$
boxes, in powers of two, provides an estimate of $D_0$.
For convenience in the generation of phase-randomized surrogates,
data sets were limited to powers of two: $2^{16}=65536$ intervals for
all but the two sleep apnea data sets, for which
$2^{14}=16384$ and $2^{13}=8192$ intervals were used respectively
(the sleep apnea data sets are much shorter than the others as
indicated in Table~A1).
For uncorrelated noise, the capacity dimension $D_0$ continues to
increase as the embedding dimension $p$ increases.
For an attractor in a deterministic system, in contrast, $D_0$
saturates as $p$ becomes larger than $2D_0 + 1$ \cite{ding93}.
Such saturation, however, is not a definitive signature of
deterministic dynamics; temporal correlation in a stochastic process
can also be responsible for what appears to be underlying
deterministic dynamics \cite{turcott96,osborne89,schiff92,theiler92}.
The distribution of the data can also play a role.

Surrogate data analysis, discussed in Sec.~\ref{sssec:surrogate},
is useful in establishing the underlying cause, by proving or
disproving various null hypotheses that the measured behavior arises
from other, non-chaotic causes.

\subsubsection{Removing Correlations in the Data}

Adjacent intervals in heartbeat sequences prove to exhibit
large degrees of correlation.
The serial correlation coefficient for RR intervals
\[
\rho (\tau) \equiv
\frac{{\rm E}[\tau_{i+1} \tau_i] - {\rm E}^2[\tau_i]}
{{\rm E}[\tau_i^2] - {\rm E}^2[\tau_i]},
\]
exceeds 0.9 for two thirds of the data sets examined, with an average
value of 0.83; the theoretical maximum is unity, with zero representing
a lack of serial correlation.
Since such large correlation is known to interfere with the detection
of deterministic dynamics, we instead employ the first difference of
the RR intervals:
$v_i \equiv \tau_i - \tau_{i + 1}$.
Such a simple transformation does not change the topological
properties of dynamical system behavior, and a related but more
involved procedure is known to reduce error in estimating other
fractal dimensions \cite{albano88}.

The serial correlation coefficient for the differenced sequence
$\{v_i\}$ turns out to have a mean value of -0.084, and none
exceeds 0.6 in magnitude.
This sequence will therefore be used to generate the embedding vector
$\stackrel{\textstyle \rightarrow}{T} \equiv
\{v_i, v_{i + 1}, ... v_{i + p-1} \}$,
with the lag $l$ set to unity since $\rho (v) \approx 0$.

\subsubsection{Surrogate Data Analysis}
\label{sssec:surrogate}

Information about the nature of an interval or segment series may be
obtained by applying various statistical measures
to surrogate data sets. These are processes constructed from the
original data in ways designed to preserve
certain characteristics of
the original data while eliminating (or modifying) others.
Surrogate data analysis provides a way of determining whether a
given
result arises from a particular property of the data set.

We make use of two kinds of surrogate data sets:
shuffled intervals and randomized phases. In
particular, we compare
statistical measures calculated from both the original data and
from its
surrogates to
distinguish those properties of the data set that arise from
correlations (such as from
long-term rate fluctuations) from those properties inherent in the
form of the original
data.

\paragraph{Shuffled intervals.}
The first class of surrogate data is
formed by shuffling (randomly reordering) the sequence of
RR intervals of the
original data set. Such random reordering destroys dependencies
among the intervals, and
therefore the correlation properties of the data, while exactly
preserving the interval histogram.
As a refinement, it is possible to divide the data into blocks and
shuffle intervals within each block, while keeping each block
separate. In this case dependencies remain over durations much
larger than the
block size, while being destroyed for durations smaller than it.

\paragraph{Randomized phases.}
The other class of surrogate data we consider is obtained by
Fourier transforming the original interval data, and then randomizing the
phases while leaving the spectral magnitudes intact.
The modified function is then inverse-transformed to return to a
time-domain representation of the intervals.
This technique exactly preserves the second-order correlation
properties of the interval sequence while removing other temporal
structure, for example that needed for phase-space reconstruction.
The interval histogram typically becomes Gaussian for this
surrogate as a result of the central limit theorem.

\subsection{Absence of Chaos}

In Fig.~10 we present plots of the capacity dimension $D_0$ plotted
against the embedding dimension $p$ for the 12 normal subjects.
Similar plots are presented in Fig.~11 for CHF patients without
atrial fibrillation, and in Fig.~12 for patients with other cardiac
pathologies.
For most of the panels in Fig.~10, the $D_0$ estimate for the
original data (solid curve) is indeed seen to rise with increasing
embedding dimension $p$, and shows evidence of approaching an
asymptotic value of about 2.
The phase-randomized data (dotted curve) yields somewhat higher
results, consistent with the presence of a deterministic attractor.

However, estimates of $D_0$ for the shuffled data (dashed curve)
nearly coincide with those of the original data.
This surrogate precisely maintains the distribution of the
relative sizes
of the differenced RR intervals, but destroys any correlation or other
dependencies in the data.
Specifically, no deterministic structure remains in the shuffled
surrogate.
Therefore, the apparent asymptote in $D_0$ must derive from the
particular form of the RR-interval distribution, and not from a
nonlinear dynamical attractor.
To quantify the closeness between results for original and shuffled
data, the shuffling process was performed ten times with ten different
random seeds.
The average of these ten independent shufflings forms the dashed
curve, with error bars representing the largest and smallest
estimates obtained for $D_0$ at each corresponding embedding
dimension.
Results for the original data indeed lie within the error bars.
Therefore, the original data does not differ from the shuffled data
at the $[(10-1)/(10+1)]100\%  = 82\%$ significance level.

Results for a good number of the other 31 data sets,
normal and pathological alike, are nearly the
same; the original data lie within the error bars of the shuffled
surrogates and consequently do not exhibit significant evidence of
deterministic dynamics. However for some records the original data
exceed the limits set by the ten shufflings.
But attributing this behavior to low-dimensional chaos
proves difficult for almost all the data records, for three reasons.
First, the surrogates often yield {\em smaller} results for $D_0$
than the original data, while shuffling data from a true chaotic
system will theoretically yield {\em larger} $D_0$ estimates.
Second, of those data sets for which the original results fall below those
of their shuffled surrogates, few achieve an asymptotic value for
$D_0$, but rather continue to climb steadily with increasing embedding
dimension.
Finally, for the remaining data sets, the original and shuffled data
yield curves that do not differ nearly as much as do those from
simulated systems known to be chaotic, such as the H\'enon attractor
(not shown) \cite{grebogi87}.
Taken together, these observations suggest that finite data length,
residual correlations, and/or other unknown effects in these
RR-interval recordings give rise to behavior in the phase-space
reconstruction that can masquerade as chaos.

Results for the phase-randomized surrogates (dotted curves in
Figs.~10-12), while exhibiting a putative capacity dimension in
excess of the original data and of the shuffled surrogates for every
patient, exhibit a remarkable similarity to each other.
This occurs because the differenced-interval sequence $\{v_i\}$
displays little correlation and the phase-randomization process
preserves this lack of correlation while imposing a Gaussian
differenced-interval distribution.
The result is nearly white Gaussian noise regardless of the original
data set, and indeed results for independently generated white
Gaussian noise closely follow these curves (not shown).

Time series measured from the heart under distinctly
non-normal operating conditions such as ventricular fibrillation
\cite{witkowski95}, or those obtained under non-physiological
conditions such as from excised hearts \cite{garfinkel95}, may exhibit
chaotic behavior under some circumstances.
This is precisely the situation
for cellular vibrations in the mammalian
cochlea \cite{heneghan94,teich95,teichchap97}. When the exciting
sound pressure level (SPL) is in the
normal physiological regime, the cellular-vibration velocities
are nonlinear but not chaotic. When the SPL is
increased beyond the normal range, however, routes to chaos emerge.
It has been suggested that since chaos-control methods prove
effective for removing such behavior, the heartbeat sequence must be chaotic.
This is an unwarranted conclusion since such methods often
work equally well for purely stochastic systems,
which cannot be chaotic \cite{christini95b}.

\section{Mathematical Models for Heart Rate Variability}
\label{sec:model}

The emphasis in this Chapter has been on HRV analysis.
The diagnostic capabilities of various measures have been presented
and compared. However the results that emerge from these studies also serve to
expose the mathematical-statistical character of the RR time series.
Such information can be of benefit in the development of
mathematical models which can be
used to simulate RR sequences. Such models may
prove useful for generating realistic heartbeat sequences in pacemakers,
for example, and for
developing improved physiological models of the cardiovascular system.
In this section we evaluate the suitability of several
integrate-and-fire
constructs for modeling heart rate variability.

\subsection{Integrate-and-Fire Model}
\label{ssec:if}

Integrate-and-fire models are widely used in the neurosciences
\cite{tuckwell89} as well as in cardiology
\cite{turcott96,berger86,hyndman73,hyndman75,rompelman82}.
These models are attractive
in part because they capture known physiology in a simple way. The
integration of a rate process can represent the cumulative effect of
neurotransmitter
on the postsynaptic membrane of a neuron, or the currents responsible
for the pacemaker potential in the sino-atrial node of the heart. The
crossing of a preset threshold by the integrated rate then gives rise
to an action potential, or a heart contraction.

The sequence of discrete events comprising the human heartbeat
can be viewed as a point process deriving from a continuous rate
function.
The integrate-and-fire method is perhaps the simplest means for
generating a point process from a rate process \cite{turcott96,thurner97}.
In this model, illustrated schematically in Fig.~13, the
rate function $\lambda(t)$ is integrated until it reaches a fixed
threshold $\theta$,
whereupon a point event is generated and the integrator is reset to zero.
Thus the occurrence time of the $(i+1)$st event is implicitly obtained from
the first occurrence of
\begin{equation}
\int_{t_{i}}^{t_{i+1}}\lambda(t)\,dt = \theta.
\label{eq:ifdef}
\end{equation}
The mean heart rate $\lambda$
(beats/sec) is given by $\lambda = {\rm E}[\lambda (t)] = 1/{\rm E}[\tau]$.
Because the conversion from the rate process
to the point process is so direct, most statistics of the point process
closely mimic those of the
underlying rate process $\lambda(t)$.
In particular, for frequencies
substantially lower than the mean heart rate, the theoretical
power spectral densities of the point process
and the underlying rate coincide \cite{thurner97}.

\subsection{Kernel of the Integrate-and-Fire Model}

The detailed statistical behavior of the point process generated by the
integrate-and-fire construct depends on the statistical character of
the rate function in the integrand of Eq.~(\ref{eq:ifdef})
\cite{turcott96}.
The fractal nature of the
heartbeat sequence requires that the rate function
$\lambda(t)$ itself be fractal.
Assuming that the underlying rate is stochastic
noise, there are three plausible candidates for $\lambda(t)$ \cite{thurner97}:
fractal Gaussian noise, fractal lognormal noise, and fractal
binomial noise. We briefly discuss these three forms of noise in turn.

\subsubsection{Fractal Gaussian Noise}
Fractal Gaussian noise served admirably in our initial efforts to
develop a suitable integrate-and-fire model for the heartbeat
sequence \cite{turcott96}.
A Gaussian process has values that, at any set of times, form a
jointly Gaussian random vector.
This property, along with its mean and spectrum, completely define
the process.
We consider a stationary rate process with a mean equal to the
expected heart rate, and a $1/f$-type {\em rate}-based spectrum that
takes the form
\cite{thurner97}
\begin{equation}
S_{\lambda}(f_{\rm time}) = \lambda^2 \delta(f_{\rm time})
+ \lambda [1 + (f_{\rm time}/f_0)^{-\alpha}].
\label{eq:fractalpsd}
\end{equation}
Here, $\lambda$ (beats/sec) and $\alpha$ are the previously defined
heart rate and fractal exponent respectively, $\delta(\cdot)$ is
again the Dirac delta function, and $f_0$ is the cutoff frequency
(cycles/sec).
Mathematically, cutoffs at low or high frequencies (or sometimes
both) are required to ensure that $\tau$ has a finite variance.
In practice, the finite duration of the simulation guarantees the
former cutoff, while the finite time resolution of the simulated
samples of $\tau$ guarantees the latter.
It is to be noted that the designation fractal Gaussian noise
properly applies only for $\alpha < 1$; the range $1 < \alpha < 3$ is
generated by fractal Brownian motion \cite{mandelbrot83}.
In the interest of simplicity, however, we employ the term fractal
Gaussian noise for all values of $\alpha$.

There are a number of recipes in the literature for generating
fractal Gaussian noise
\cite{flandrin92,mandelbrot71,lundahl86,stoksik94}.
All typically result in a sampled version of fractal Gaussian noise,
with equal spacing between samples.
A continuous version of the signal is obtained by using a
zeroth-order interpolation between the samples.

The use of a fractal Gaussian noise kernel in the
integrate-and-fire construct
results in the so-called fractal-Gaussian-noise driven
integrate-and-fire model \cite{turcott96,thurner97}.
The mean of the rate process is generally required to be
much larger than its standard
deviation, in part so that the times when the rate is negative
(during which
no events may be generated) remain small.

The {\em interval}-based spectrum is obtained by applying
Eq.~(\ref{eq:freqconv}) to Eq.~(\ref{eq:fractalpsd}) (see
Sec.~\ref{ssec:stdfreqm} and Ref.~\cite{deboer84}).
The following approximate result, in terms of interval-based
frequency $f$, is obtained:
\begin{equation}
S_{\tau}(f) \approx \lambda^{-2} [\delta(f) + 1 + (\lambda f/f_0)^{-\alpha}].
\end{equation}

We proceed to discuss two other physiologically based
noise inputs \cite{thurner97}; both reduce to fractal Gaussian
noise in appropriate limits.

\subsubsection{Fractal Lognormal Noise}

A related process results from passing fractal Gaussian noise through
a memoryless exponential transform.
This process plays a role in neurotransmitter exocytosis
\cite{sblpoo}.
Since the exponential of a Gaussian is lognormal, we refer to this
process as fractal lognormal noise \cite{thurner97,sblpoo}.
If $X(t)$ denotes a fractal Gaussian noise process with mean ${\rm
E}[X]$, variance ${\rm Var}[X]$, and autocovariance function
$K_X(\tau)$, then $\lambda(t) \equiv \exp[X(t)]$ is a fractal
lognormal noise process with moments
${\rm E}[\lambda^n] = \exp(n{\rm E}[X] + n^2{\rm Var}[X]/2)$
and autocovariance function
$K_\lambda(\tau) = {\rm E}^2[\lambda](\exp[K_X(\tau)] - 1)$
\cite{sblpoo}.

By virtue of the exponential transform, the autocorrelation functions
of the Gaussian process $X(t)$ and the lognormal process $\lambda(t)$
differ; thus their spectra differ. In particular, the power spectral
density of the resulting point process does not follow the exact form
of Eq.~(\ref{eq:fractalpsd}), although for small values of
${\rm Var}[X]$ the experimental periodogram closely
resembles this ideal form~\cite{sblpoo}.
In the limit of very small values of ${\rm Var}[X]$, the exponential
operation approaches a linear transform, and the rate process reduces
to fractal Gaussian noise.

\subsubsection{Fractal Binomial Noise}

A third possible kernel for the integrate-and-fire model is fractal
binomial noise.
This process results from the addition of a number of independent
identical alternating fractal renewal processes.
The sum is a binomial process with the same fractal exponent as each
of the individual alternating fractal processes
\cite{lowen93b,lowen93c}.
This binomial process can serve as a rate function for an
integrate-and-fire process; the result is the fractal-binomial-noise
driven integrate-and-fire model \cite{thurner97}.
This construct was initially designed to model the superposition of
alternating currents from a number of ion channels with fractal
behavior \cite{lowen93b,lowen93c,lowen93a,liebovitch90}.
As the
number of constituent processes increases, fractal binomial noise
converges to fractal Gaussian noise with the same fractal exponent
therefore leading to
the fractal-Gaussian-noise driven integrate-and-fire
point process \cite{thurner97}.

\subsection{Jittered Integrate-and-Fire Model}
\label{ssec:jif_model}

Conventional integrate-and-fire constructs have only a single source
of randomness, the rate process $\lambda(t)$.
It is sometimes useful to incorporate a second source of randomness
into such models.
One way to carry this out is to impose random jitter on the
interevent times generated in the integrate-and-fire process
\cite{thurner97}.

The procedure used for imposing this jitter is as follows.
After generating the time of the $(i+1)$st event $t_{i+1}$ in
accordance with Eq.~(\ref{eq:ifdef}), the $(i+1)$st interevent time
$\tau_{i+1} =  t_{i+1} - t_i$ is multiplied by a dimensionless
Gaussian-distributed random variable with unit mean and variance
$\sigma^2$.
This results in the replacement of $t_{i+1}$ by
\begin{equation}
t_{i+1} + \sigma (t_{i+1} - t_i) {\cal N}_i
\label{eq:tupdate}
\end{equation}
before the simulation of subsequent events
($t_{i+2}$, $t_{i+3}$,\ldots).
The quantity $\{{\cal N}_i\}$ represents a sequence of zero-mean,
unity-variance independent Gaussian
random variables and the
standard deviation $\sigma$ is a free parameter that controls the
strength of the
proportional jitter. In accordance with the construction of the model,
events at subsequent times experience jitter imposed
by all previous events (see Fig.~14).

The overall result is the jittered integrate-and-fire model.
The jitter serves to introduce
a source of scale-independent noise into the point process.
Depending on
the character of the
input rate function, this model can give rise to the fractal-Gaussian-noise,
fractal-lognormal-noise, or fractal-binomial-noise
driven jittered integrate-and-fire
processes, among others.
Two limiting behaviors readily emerge from this model. When
$\sigma \to 0$,
the jitter disappears and the jittered integrate-and-fire
model reduces to the ordinary integrate-and-fire model.
At the other extreme, when $\sigma \to \infty$, the jitter dominates
and the result is a homogeneous Poisson process.
The fractal behavior present in the rate function $\lambda(t)$ then
disappears and $\lambda(t)$ behaves as if it were constant.
Between these two limits, as $\sigma$ increases the fractal onset
time $1/f_0$ of the resulting point process increases and the fractal
characteristics of the point process are progressively lost, first at
higher frequencies (shorter times) and subsequently at lower
frequencies (longer times) \cite{thurner97}.

\subsubsection{Simulating the Jittered Integrate-and-Fire Point Process}

We use a fractal Gaussian noise kernel in the jittered
integrate-and-fire construct to simulate the human heartbeat.
Fractal Gaussian noise with the appropriate mean, standard deviation,
and fractal exponent $\alpha_S$ is generated using Fourier-transform
methods.
This noise serves as the kernel in an integrate-and-fire element
[Eq.~(\ref{eq:ifdef})], the output of which is jittered in accordance
with Eq.~(\ref{eq:tupdate}).
For each of the 12 normal subjects and 15 CHF ($\bar{\rm s}$ and
$\bar{\rm c}$ AF) patients, the model parameters were adjusted
iteratively to obtain a simulated data set that matched the patient
recording as closely as possible.
In particular, each of the 27 simulations contained exactly $L_{\rm
max}=75821$ RR intervals, displayed mean heart rates that fell within
1\% of the corresponding data recordings, and exhibited
wavelet-transform standard deviations that were nearly identical with
those of the data.
The simulation results for CHF patients with and without atrial
fibrillation were very similar.

\subsubsection{Statistics of the Simulated Point Process for Normal
Subjects and CHF Patients}

Having developed the jittered integrate-and-fire model and simulated
sequences of RR intervals, we proceed to examine some of the statistical
features of the resulting point processes.

Figure 15 displays representative results in which the data (solid
curves) and simulations (dotted curves) are compared for a single
normal subject (left column) and a single CHF ($\bar{\rm s}$ AF)
patient (right column).
Figure~15(a) illustrates that the RR-interval sequences obtained from
the model and the data are qualitatively similar, for both the normal
and heart-failure cases.
Furthermore, the agreement between the simulated and actual
wavelet-transform standard-deviation curves is excellent in both
cases, as is evident in Fig.~15(b).
Even the gentle increase of $\sigma_{\rm wav}$ for the heart-failure
patient at small scale values is reproduced, by virtue of the jitter
introduced into the model.
It is apparent in Fig.~15(c), however, that the simulated RR-interval
histogram for the heart-failure patient does not fit the actual
histogram well; it is too broad which indicates that $\sigma_{\rm
int}$ of the simulation is too large.
Finally, Fig.~15(d) shows that the simulated spectra of the RR
intervals match the experimental spectra quite nicely, including the
whitening of the heart-failure spectrum at high frequencies.
This latter effect is the counterpart of the flattening of
$\sigma_{\rm wav}$ at small scales, and results from the white
noise introduced by the jitter at high frequencies.

\subsubsection{Simulated Individual Value Plots and ROC-Area Curves}

To examine how well the jittered integrate-and-fire model mimics the
collection of data from a global perspective, we constructed
individual-value plots and ROC-area curves using
the 27 simulated data sets.
The results are presented in Figs.~16 and~17 respectively.
Direct comparison should be made with Figs.~5 and~8, respectively,
which provide the same plots for the actual heart-failure patient and
normal-subject data.

Comparing Fig.~16 with Fig.~5, and Fig.~17 with Fig.~8, we see that in
most cases the simulated and actual results are remarkably
similar. Based on their overall ability to distinguish
between CHF and normal simulations for the full set of $75821$ RR
intervals,
the 16 measures portrayed in Fig.~16 roughly fall
into three classes. A similar division was observed for the actual
data, as discussed in Sec.~\ref{ssec:compmeas}

As is evident in Fig.~16, five measures (indicated by boldface font)
succeed in fully separating the two kinds of simulations and, by
definition, fall into the first class:
VLF, LF, $\sigma_{\rm wav}(32)$, $S_{\tau}(1/32)$, and DFA(32).
These measures share a dependence on a single scale, or small range
of scales, near 32 heartbeat intervals.
However, of these five measures only two (LF and VLF) successfully
separate the simulated results for the next smaller number of RR
intervals ($L=65536$) and none succeeds in doing so for yet smaller
values of $L$.

These same five measures also fully distinguish the normal subjects
from the CHF patients (see Fig.~5); however a sixth measure, $A(10)$,
also achieves this.
For the actual data all six of these measures successfully separate
the actual heart-failure patients from the normal subjects for data
lengths down to $L=32768$ RR intervals.

The simulated ROC-area curves presented in Fig.~17 resemble the
curves for the normal and CHF data shown in Fig.~8, but there are a
few notable distinctions.
The simulated and actual ROC areas (Figs.~17 and~8 respectively) for
the interval measures SDANN and $\sigma_{\rm int}$ are just about the
same for $L=64$ RR intervals.
However both of these measures exhibit {\em decreasing} ROC areas as
the number of simulated RR intervals ($L$) increases (Fig.~17).
In contrast, the ROC areas based on the actual data increase with
increasing data length $L$ in normal fashion (Fig.~8).
The decrease in ROC area observed in Fig.~17 means that performance
is degraded as the simulated data length increases.
This paradoxical result likely arises from a deficiency of the model;
it proved impossible to simultaneously fit the wavelet-transform
standard deviation at all scales, the mean heart rate, and the
RR-interval standard deviation.
Choosing to closely fit the two former quantities to the data
rendered inaccurate the simulated interval standard deviation.
Since shorter files have insufficient length to exhibit the longer
term fluctuations that dominate fractal activity, the interval
standard deviation for small values of $L$ is not influenced by these
fluctuations.
Apparently the manner in which the long-term effects are expressed
differs in the original data and in the simulation.

The most surprising distinction between the actual and simulated
results, perhaps, is the dramatic putative
improvement in the ability of the
five scale-independent fractal-exponent measures to separate simulated
heart-failure and normal data, as the number of RR intervals
increases. The Allan factor exponent $\alpha_A$ appears to
offer the greatest ``improvement'', as is seen by comparing
Figs.~17 and~8.
All of the exponents except $\alpha_W$ attain an ROC area
exceeding 0.97 for $L_{\rm max}=75821$ RR intervals. Nevertheless,
the improved separation yields performance that remains
inferior to that of the fixed-scale
measures. The apparent improved separation in the simulated
data may therefore
arise from a nonlinear interaction in the model between the fractal
behavior and the signal component near a scale of 32 heartbeat intervals.
Or it may be that
some aspect of the data, not reliably detected by fractal-exponent
measures applied to the original data, emerges more clearly following
the modeling and simulation. This could mean that
another measure of the
fractal exponent of the actual data, perhaps incorporating
in some way the processing provided by the
modeling procedure, might yield better results than the more
straightforward fractal-exponent measures that we have used to this point.
 
\subsubsection{Limitations of the Jittered Integrate-and-Fire Model}
\label{sssec:jiflims}

Figure~18 presents a direct comparison between four measures derived
from the original data and from the corresponding simulations.
Each panel contains 12 normal results ($+$), 12 CHF $\bar{\rm{s}}$ AF
results ($\times$), and 3 CHF $\bar{\rm{c}}$ AF results
($\triangle$).
The diagonal lines correspond to perfect agreement.

The results for $S_{\tau}(1/32)$ and $\sigma_{\rm wav}(32)$ are
excellent, illustrating the remarkable ability of the model to mimic
the data at a time scale of 32 interbeat intervals.
The correlation coefficients for these two measures lie very close to
unity, indicating almost perfect correspondence between the original
and simulated data.

The upper left panel presents values for $\sigma_{\rm int}$, the
standard deviation of the RR intervals.
Agreement between the simulation and original data is only fair, with
simulated values of $\sigma_{\rm int}$ generally somewhat higher than
desired.
The correlation coefficient $\rho=0.71$ for all 27 simulations
indicates significant correlation ($p<0.0003$ that 27 random pairs of
numbers with the same distribution would achieve a magnitude of $\rho
\ge 0.71$), but not extremely close agreement between simulated and
original RR-interval sequences.

This disagreement highlights a problem with the simple jittered
integrate-and-fire model. The fractal-Gaussian-noise integrate-and-fire
model {\em without} jitter yields substantially
closer agreement between the data and simulation for this statistic
\cite{turcott96}. Although the introduction of
jitter leads to improved agreement for
$\sigma_{\rm wav}$, it brings the model results further away from
the data as measured by
$\sigma_{\rm int}$. Apparently,
a method other than jitter must be devised to
forge increased agreement with
$\sigma_{\rm wav}$, while not degrading the agreement with
$\sigma_{\rm int}$. One possibility is the reordering of
the individual RR intervals over short
time scales.

The simulated exponent $\alpha_W$ also fails to show close
agreement with the data, suggesting that there are other
features of the model that are less than ideal.
However, since this measure is of little use in separating CHF
patients from normal subjects, this disagreement takes on reduced
importance.

\subsection{Toward an Improved Model of Heart Rate Variability}

Agreement between the simulations and data
would most likely be improved by adding other inputs
to the model aside from fractal noise, as illustrated
schematically in Fig.~19. Input signals related to
respiration and blood pressure are two likely candidates
since it is well known that variations in these physiological functions
influence heart rate variability.
Experience also teaches us that there is a
measure of white Gaussian noise present,
possibly associated with
the measurement process.

As indicated in Sec.~\ref{sssec:jiflims}, the introduction of jitter
in the integrate-and-fire construct improves the agreement of the
model with certain features of the the data, but degrades it for
others.
An adaptive reset will provide a more flexible alternative to the
jitter.
Moreover, the threshold $\theta$ could be converted into a stochastic
process as a way of incorporating variability in other elements of
the system \cite{thurner97}.

All things considered, the jittered integrate-and-fire construct does
a rather good job of mimicking the actual data, though the
introduction of a more physiologically based model would be a welcome
addition.

\section{Conclusion}

For the purposes of heart-rate-variability analysis, the occurrence
times of the sequence of R phases in the electrocardiogram can be
represented as a fractal-rate stochastic point process.
Using a collection of standard and novel heart-rate-variability
measures, we have examined the sequence of times between the R
phases, {\em viz.} the RR time series, of ECGs recorded from normal
subjects and congestive heart-failure patients, as well as from
several patients with other cardiovascular disorders.
Congestive-heart-failure patients who also suffered from atrial
fibrillation were treated as a separate class.
We examined sixteen heart-rate-variability measures, comprising
frequency-domain, time-domain, scaling-exponent, Allan-factor,
detrended-fluctuation-analysis, and wavelet-transform methods.

Receiver-operating-characteristic analysis was used to compare and
contrast the performance of these sixteen measures with respect to
their abilities to properly distinguish heart-failure patients from
normal subjects over a broad range of record lengths (minutes to
hours).
Scale-dependent measures rendered performance that was substantially
superior to that of scale-independent ones.
Judged on the basis of performance and computation time, two measures
of the sixteen emerged at the top of the list: the wavelet-transform
standard deviation $\sigma_{\rm wav}(32)$ and the RR-interval-based
power spectral density $S_{\tau}(1/32)$.
They share in common the dependence on a single scale, or small range
of scales, near 32 heartbeat intervals.
The behavior of the ECG at this particular scale, corresponding to
about 25 sec, turns out to be a significant marker for the presence
of cardiac dysfunction.
The scale value itself is far more important than the details of the
measures used to examine it.

Application of these techniques to a large database of normal and CHF
records will make it possible to uncover just how the ECGs of
pathological patients differ from those of normal subjects in the
vicinity of this special scale.
This would facilitate the development of an optimal diagnostic
measure, which might take the form of a specialized weighting
function over the interval-based power spectral density or over some
combination of elementary measures.

We also addressed an issue of fundamental importance in cardiac
physiology: the determination of whether the RR time series arises
from a chaotic attractor or has an underlying stochastic origin.
Using nonlinear-dynamics theory, together with surrogate-data
analysis, we established that the RR sequences from both normal
subjects and pathological patients have stochastic, rather than
deterministic, origins.
The use of a special embedding of the {\em differences} between
adjacent RR intervals enabled us to reach this conclusion.
This technique substantially reduced the natural correlations
inherent in the interbeat-interval time series which can masquerade
as chaos and confound simpler analyses.

Finally, we developed a jittered integrate-and-fire model built
around a fractal-Gaussian-noise kernel.
Simulations based on this model provided realistic, though not
perfect, replicas of real heartbeat sequences.
The model was least successful in predicting interbeat-interval
statistics and fractal-exponent values.
Nevertheless it could find use in some applications such as pacemaker
excitation.

To confirm the observations we have reported, it will be desirable to
carry out continuation studies using large numbers of out-of-sample
records.
It will also be useful to examine how the performance of the various
measures correlates with the severity of cardiac dysfunction as
judged, for example, by the NYHA clinical scale.
Comparisons with existing CHF studies that make use of HRV measures
should also be conducted.

\pagebreak

\section*{Appendix A}
\addcontentsline{toc}{section}{Appendix A}

Table~A1 provides a summary of some significant statistics of the
RR records analyzed in this Chapter, before truncation to $75821$
intervals.

\pagebreak

\addcontentsline{toc}{section}{References}
\bibliography{lowen}
\bibliographystyle{ieeetr}

\pagebreak

\section*{Figure Captions}
\addcontentsline{toc}{section}{Figure Captions}

Fig.~1. In HRV analysis the electrocardiogram (ECG) schematized in
(a) is represented by a sequence of times of the R phases which form
an unmarked point process [vertical arrows in (b)]. This sequence may
be analyzed as a sequence of counts $\{N_i\}(T)$ in a predetermined
time interval $T$ as shown in (c), or as a sequence of interbeat (RR)
intervals $\{\tau_i\}$ as shown in (d). The sequence of counts forms
a discrete-time random counting process of nonnegative integers
whereas the sequence of intervals forms a sequence of positive
real-valued random numbers.

Fig.~2. Estimating the wavelet transform using the Haar wavelet:
(a) original Haar wavelet; (b) delayed and scaled
version of the wavelet ($m=16, n=3$); and (c) time series multiplied
by this wavelet.

Fig.~3.
(a) Series of interbeat intervals $\tau_i$ versus  interval number
$i$ for a typical normal patient (data set n16265).
(Adjacent values of the interbeat interval are connected by straight
lines to facilitate viewing.)
Substantial trends are evident.
The interbeat-interval standard deviation $\sigma_{\rm int}\equiv $
SDNN is indicated.
(b) Wavelet transform $W_{m,n}(m)$ (calculated using a Daubechies
10-tap analyzing wavelet) at three scales ($m$ = 4, 16, 256)
vs.\ interval number $i$ (=$mn$) for the RR-interval sequence shown
in (a).
The trends in the original interbeat-interval time series are removed
by the wavelet transformation.
The wavelet-transform standard deviation $\sigma_{\rm wav}(m)$ for
this data set is seen to increase with the scale $m$.

Fig.~4.
Haar-wavelet-transform standard deviation $\sigma_{\rm wav}(m)$ vs
scale $m$ for the 12 normal subjects ($+$), 12 CHF patients without
($\bar{\rm{s}}$) atrial fibrillation ($\times$), and the 3 CHF
patients with ($\bar{\rm{c}}$) atrial fibrillation ($\triangle$).
Each data set comprises the first $75821$ RR intervals of a recording
drawn from the Beth-Israel Hospital heart-failure database.
Complete separation of the normal subjects and heart-failure patients
is achieved at scales $m=16$ and 32 interbeat intervals.

Fig.~5.  Individual value plots (data) for the 16 measures.
Each panel corresponds to a different HRV measure.
The seven columns in each panel, from left to right, comprise data
for
(1) 12 normal subjects ($+$),
(2) 12 CHF patients $\bar{\rm{s}}$ AF ($\times$),
(3) 3 CHF patients $\bar{\rm{c}}$ AF ($\triangle$),
(4) 1 heart-transplant patient (\hbox{$\times$ \hspace{-1.45em} $+$}),
(5) 1 ventricular tachycardia patient ($\diamondsuit$),
(6) 1 sudden-cardiac-death patient
(\hbox{$\sqcup$ \hspace{-1.35em} $\sqcap$}), and
(7) 2 sleep apnea patients ($+$).
Each data set comprises $75821$ RR intervals except for the two sleep
apnea data sets which comprise $16874$ and $15751$ RR intervals
respectively.
The six measures highlighted in boldface font succeed in completely
separating normal subjects and CHF patients
($\bar{\rm {s}}$ and $\bar{\rm {c}}$ atrial fibrillation)
{\boldmath {\bf VLF}, {\bf LF}, $A(10)$, $\sigma_{\rm wav}(32)$,
$S_{\tau}(1/32)$, and {\bf DFA(32)}.}

Fig.~6. Positive (solid curves) and negative (dotted curves)
predictive values for all 16 HRV measures plotted against the
threshold $\theta$, each in its own panel.
These curves are constructed using the 12 normal and 12 heart-failure
($\bar{\rm{s}}$ AF) records drawn from the CHF database, each of
which has been truncated to $75821$ RR intervals.
The six measures highlighted in boldface font exhibit threshold
regions for which both the positive and negative predictive values
are unity:
{\boldmath {\bf VLF}, {\bf LF}, $A(10)$, $\sigma_{\rm wav}(32)$,
$S_{\tau}(1/32)$, and {\bf DFA(32)}.}
This indicates that the normal subjects and CHF ($\bar{\rm{s}}$ AF)
patients can be completely distinguished by these six measures, in
accordance with the results established in Fig.~5.

Fig.~7. Receiver-operating-characteristic (ROC) curves (sensitivity
vs $1 - \rm {specificity}$) for all 16 HRV measures, each in its own
panel.
These curves are constructed using the 12 normal and 12 heart-failure
($\bar{\rm{s}}$ AF) records drawn from the CHF database, each of
which has been truncated to $75821$ RR intervals.
The six measures highlighted in boldface font simultaneously achieve
100\% sensitivity and 100\% specificity so that the ROC curve is
perfectly square:
{\boldmath {\bf VLF}, {\bf LF}, $A(10)$, $\sigma_{\rm wav}(32)$,
$S_{\tau}(1/32)$, and {\bf DFA(32)}.}
This indicates that the normal subjects and CHF ($\bar{\rm{s}}$ AF)
patients can be completely distinguished by these six measures, in
accordance with the results established in Figs.~5 and~6.

Fig.~8. Diagnostic accuracy (area under ROC curve) as a function of
record length analyzed $L$ (number of RR intervals) for the 16 HRV
measures (mean $\pm 1$ S.D.).
An area of unity corresponds to perfect separability of the two
classes of patients.
The six measures highlighted in boldface font
({\boldmath {\bf VLF}, {\bf LF}, $A(10)$, $\sigma_{\rm wav}(32)$,
$S_{\tau}(1/32)$, and {\bf DFA(32)} })
provide such perfect separability at the longest record lengths, in
accordance with the results in Fig.~7.
As the record length decreases performance degrades at a different
rate for each measure.
The five scale-independent measures,
$\alpha_D$, $\alpha_W$, $\alpha_S$, $\alpha_A$, and $\alpha_R$,
perform poorly at all data lengths.

Fig.~9.  $1-$ROC area as a function of data length $L$, on
a doubly logarithmic plot, for the six most effective measures:
VLF, $A(10)$, DFA(32),
$\sigma_{\rm wav}(32)$, $S_{\tau}(1/32)$, and LF.
The unconventional form of this ROC plot
allows small
deviations from unity area to be highlighted.
The trends exhibited by all six measures are broadly
similar, but
VLF, $A(10)$, and DFA(32) exhibit more degradation
at shorter data lengths than do the other three measures.
Thus three measures emerge as superior: {\boldmath
{\bf LF}, $\sigma_{\rm wav}(32)$, and $S_{\tau}(1/32)$}. When
computation time is taken into account in accordance with
the results provided in Table~2,
{\boldmath $\sigma_{\rm wav}(32)$
and $S_{\tau}(1/32)$} are the two preferred measures.

Fig.~10. Capacity dimension $D_0$ as a function of
embedding dimension $p$ for the 12 normal subjects. The
three curves
shown for each subject correspond to
the original differenced intervals (solid curves),
the shuffled surrogates
(dashed curves), and the phase-randomized surrogates (dotted curves).
The dashed curves are quite similar to the solid curves in each
panel while the 12 dotted
curves closely resemble each other.

Fig.~11. Capacity dimension $D_0$ as a function of
embedding dimension $p$ for the 12 CHF patients without atrial
fibrillation. The
three curves
shown for each patient correspond to
the original differenced intervals (solid curves),
the shuffled surrogates
(dashed curves), and the phase-randomized surrogates (dotted curves).
The dashed curves are reasonably similar to the solid curves in most
panels while the 12 dotted
curves closely resemble each other.

Fig.~12. Capacity dimension $D_0$ as a function of
embedding dimension $p$ for the eight patients with other
cardiac pathologies. The
three curves
shown for each patient correspond to
the original differenced intervals (solid curves),
the shuffled surrogates
(dashed curves), and the phase-randomized surrogates (dotted curves).
The dashed curves are reasonably similar to the solid curves in most
panels while the 8 dotted
curves closely resemble each other.

Fig.~13. Simple integrate-and-fire model often
used in cardiology and neurophysiology. A rate function
$\lambda (t)$ is integrated until it reaches a preset
threshold $\theta$ whereupon a point event is generated
and the integrator is reset to zero.

Fig.~14. Schematic representation of the jittered
integrate-and-fire model for generating a simulated
RR-interval series. An
underlying rate process $\lambda (t)$, assumed to be bandlimited
fractal Gaussian noise, is integrated until it reaches a fixed threshold
$\theta$, whereupon a point event is generated. The occurrence
time of the point event is jittered
in accordance with a Gaussian distribution and the
integrator is then reset. The continuous
rate process is thereby converted
into a point process representing
the sequence of R phases in the human heartbeat.

Fig.~15. Comparison between data (solid curves) and simulations
(dotted curves) for a single normal subject (left column, data set
n16265), and a single CHF $\bar{\rm s}$ AF patient (right column,
data set a6796).
The parameters ${\rm E}[\tau_i]$ (representing the mean interbeat
interval) and $\alpha_W$ (representing the fractal exponent) used in
the simuluations were derived from the data for the two individuals.
The normal-subject simulation parameters were $1/\lambda=0.74$ sec,
$\alpha_S=1.0$, $f_0=0.0012$ cycles/sec, and $\sigma=0.022$;
the CHF-patient simulation parameters were $1/\lambda=0.99$ sec,
$\alpha_S=1.5$, $f_0=0.00055$ cycles/sec, and $\sigma=0.025$.
The lengths of the simulated data sets were identical to those of the
experimental records analyzed.
(a) RR-interval sequence over the entire data set.
Qualitative agreement is apparent in both the normal and
heart-failure panels.
(b) Wavelet-transform standard deviation versus scale.
The simulations mimic the scaling properties of the data in both
cases, as well as the gentle flattening of $\sigma_{\rm wav}$ for the
heart-failure patient at small scale values.
(c) Interbeat-interval histogram.
The model is satisfactory for the normal subject but fails to capture
the narrowing of the histogram (reduction of $\sigma_{\rm int}$) for
the heart-failure patient.
(d) Spectrum of the sequence of RR intervals (the data are displaced
upward by a factor of $10^2$ for clarity).
The simulations capture the subtleties in the spectra quite well,
including the whitening of the heart-failure spectrum at high
frequencies.

Fig.~16. Individual value plots (jittered
integrate-and-fire simulations) for the 16 measures.
Each panel corresponds to a different HRV measure.
The three columns in each panel, from left to right, comprise data
for
(1) 12 simulations using normal-subject parameters ($+$),
(2) 12 simulations using CHF-patient ($\bar{\rm{s}}$ AF) parameters
($\times$), and
(3) 3 simulations using CHF-patient ($\bar{\rm{c}}$ AF) parameters
($\triangle$).
Each simulation comprises $75821$ RR intervals and is carried
out using parameters drawn from a single actual data set.
The five measures highlighted in boldface font succeed in completely
separating normal-subject and CHF-patient
($\bar{\rm {s}}$ and $\bar{\rm {c}}$ atrial fibrillation) simulations:
{\boldmath {\bf VLF}, {\bf LF}, $\sigma_{\rm wav}(32)$,
$S_{\tau}(1/32)$, and {\bf DFA(32)}.}
Each panel should be compared with the corresponding panel
for the actual normal and CHF data in Fig.~5 (leftmost three columns).

Fig.~17. Area under the ROC curve (jittered integrate-and-fire
simulations) as a function of number of RR intervals analyzed ($L$)
for the 16 HRV measures (mean $\pm 1$ S.D.).
An area of unity corresponds to perfect separability of the two
classes of simulations.
The five measures highlighted in boldface font
[{\boldmath {\bf VLF}, {\bf LF}, $\sigma_{\rm wav}(32)$,
$S_{\tau}(1/32)$, and {\bf DFA(32)}}]
provide such perfect separability, but only for the largest number of
RR intervals analyzed.
Each panel should be compared with the corresponding panel for the
actual normal and CHF data in Fig.~8.
The ROC-area simulations differ from those for the data in two
principal respects: the performance of the two interval measures
SDANN and $\sigma_{\rm int}$ severely degrades as the number of RR
intervals increases, and the performance of the five fractal-exponent
measures is substantially enhanced as the number of RR intervals
increases.

Fig.~18.  Simulation accuracy for four measures with their
correlation coefficients $\rho$. The jittered
integrate-and-fire model performs
remarkably well for the measures $S_{\tau}(1/32)$ and
$\sigma_{\rm wav}(32)$; however it does not perform nearly as well
for the measures
$\sigma_{\rm int}$ and $\alpha_W$. These disagreements highlight
problems with the simple jittered
integrate-and-fire model.

Fig.~19. Potential modifications to the simple integrate-and-fire model.
Multiple
physiologically based inputs and adaptive feedback could serve to
produce more realistic RR-interval simulations.

\pagebreak

\section*{Tables}
\addcontentsline{toc}{section}{Tables}

\noindent
TABLE 1:
Detection distances $h$ and $d$ for all 16 HRV
measures applied to the 12 normal and 12 CHF records
of length $L_{\rm max}$.
The measures are listed in order of descending value of $h$.
The rankings according to $d$ differ from those according to $h$.
The wavelet-transform standard deviation $\sigma_{\rm wav}(32)$ and
variance $\sigma_{\rm wav}^2(32)$, though related by a simple
monotonic transformation, yield different values of $h$ and
have different rankings.
\begin{center}
\begin{tabular}{lll}
Measure                  & \multicolumn{1}{c}{$h$} &
   \multicolumn{1}{c}{$d$}\\
\hline
DFA$(32)$                & 2.48253 & 1.81831 \\
$\sigma_{\rm wav}(32)$   & 2.33614 & 1.70153 \\
$A(10)$                  & 2.32522 & 1.77482 \\
VLF                      & 1.84285 & 1.56551 \\
$S_\tau(1/32)$           & 1.77422 & 1.55200 \\
$\sigma_{\rm int}$       & 1.74750 & 1.32475 \\
$\sigma_{\rm wav}^2(32)$ & 1.71343 & 1.47165 \\
$\alpha_D$               & 1.64883 & 1.17679 \\
SDANN                    & 1.46943 & 1.04079 \\
LF                       & 1.36580 & 1.28686 \\
pNN50                    & 1.36476 & 1.20896 \\
LF$/$HF                  & 1.24507 & 0.91444 \\
$\alpha_W$               & 1.09916 & 0.77800 \\
$\alpha_R$               & 1.02367 & 0.72463 \\
HF                       & 0.85361 & 0.73077 \\
$\alpha_S$               & 0.82125 & 0.58071 \\
$\alpha_A$               & 0.38778 & 0.27895 \\
\end{tabular}
\end{center}

\pagebreak

\noindent
TABLE 2: Computation times (to the nearest 10 msec) for the
16 HRV measures
for data sets comprising
$75821$ RR intervals.
The long execution times for the two DFA measures results
from the fact that it is an $N^2$
process whereas the 14 other methods are either $N$ or $N \log(N)$.
\begin{center}
\begin{tabular}{lr} & Execution\\ Measure & Time (msec)\\
\hline
VLF, LF, HF, and LF$/$HF & 330\\
pNN50 & 40\\
SDANN & 160\\
$\sigma_{\rm int}$ & 190\\
$A(10)$ & 160\\
$\sigma_{\rm wav}(32)$ & 20\\
$S_\tau(1/32)$ & 60\\
DFA$(32)$ & 650,090\\
$\alpha_D$ & 650,110\\
$\alpha_W$ & 220\\
$\alpha_S$ & 920\\
$\alpha_A$ & 610\\
$\alpha_R$ & 570\\
\end{tabular}
\end{center}

\pagebreak

\noindent
TABLE A1: Elementary
statistics of the original RR-interval recordings, before
truncation to $75821$ RR intervals.
\begin{center}
\begin{tabular}{l@{\hspace{-0.2mm}}c@{\hspace{-4.2mm}}rcccc}
File- & & \multicolumn{1}{c}{Number of} & Recording & RR-Interval &
Mean Rate & RR-Interval\\
name & Condition & \multicolumn{1}{c}{RR Intervals} & Dur.\ (sec) &
Mean (sec) & (sec)$^{-1}$ & S.D.\ (sec)\\
\hline
n16265 & Normal & 100460 & 80061 & 0.797 & 1.255 & 0.171\\
n16272 & " & 93177 & 84396 & 0.906 & 1.104 & 0.142\\
n16273 & " & 89846 & 74348 & 0.828 & 1.208 & 0.146\\
n16420 & " & 102081 & 77761 & 0.762 & 1.313 & 0.101\\
n16483 & " & 104338 & 76099 & 0.729 & 1.371 & 0.089\\
n16539 & " & 108331 & 84669 & 0.782 & 1.279 & 0.150\\
n16773 & " & 82160 & 78141 & 0.951 & 1.051 & 0.245\\
n16786 & " & 101630 & 84052 & 0.827 & 1.209 & 0.116\\
n16795 & " & 87061 & 74735 & 0.858 & 1.165 & 0.212\\
n17052 & " & 87548 & 76400 & 0.873 & 1.146 & 0.159\\
n17453 & " & 100674 & 74482 & 0.740 & 1.352 & 0.103\\
nc4 & " & 88140 & 71396 & 0.810 & 1.235 & 0.132\\
a6796 & HF $\bar {\rm s}$ AF & 75821 & 71934 & 0.949 & 1.054 & 0.091\\
a8519 & " & 80878 & 71948 & 0.890 & 1.124 & 0.062\\
a8679 & " & 119094 & 71180 & 0.598 & 1.673 & 0.051\\
a9049 & " & 92497 & 71959 & 0.778 & 1.285 & 0.058\\
a9377 & " & 90644 & 71952 & 0.794 & 1.260 & 0.060\\
a9435 & " & 114959 & 71158 & 0.619 & 1.616 & 0.034\\
a9643 & " & 148111 & 71958 & 0.486 & 2.058 & 0.024\\
a9674 & " & 115542 & 71968 & 0.623 & 1.605 & 0.084\\
a9706 & " & 115064 & 71339 & 0.620 & 1.613 & 0.100\\
a9723 & " & 115597 & 71956 & 0.622 & 1.607 & 0.027\\
a9778 & " & 93607 & 71955 & 0.769 & 1.301 & 0.070\\
a9837 & " & 115205 & 71944 & 0.624 & 1.601 & 0.066\\
a7257 & HF $\bar {\rm c}$ AF & 118376 & 71140 & 0.601 & 1.664 & 0.039\\
a8552 & " & 111826 & 71833 & 0.642 & 1.557 & 0.062\\
a8988 & " & 118058 & 71131 & 0.603 & 1.660 & 0.091\\
tp987 & TRANSPLANT & 106394 & 67217 & 0.632 & 1.583 & 0.083\\
vt973 & VT & 86992 & 70044 & 0.805 & 1.242 & 0.202\\
sd984  &  SCD & 77511 & 62786 & 0.810 & 1.234 & 0.089\\
slp59 & APNEA & 16874 & 14399 & 0.853 & 1.172 & 0.103\\
slp66 & " & 15751 & 13199 & 0.838 & 1.193 & 0.103\\
\end{tabular}
\end{center}

\pagebreak

\section*{Contact Information}
\addcontentsline{toc}{section}{Contact Information}

\noindent
Malvin~C.~Teich \\
Departments of Electrical and Computer Engineering,
Biomedical Engineering, and Physics, \\
Boston University, 8 Saint Mary's Street, Boston, Massachusetts 02215-2421, USA \\
Phone: 617-353-1236;
Fax: 617-353-1459 \\
E-mail: teich@bu.edu;
URL: http://people.bu.edu/teich/

\noindent
Steven~B.~Lowen \\
Department of Psychiatry, Harvard Medical School, \\
Developmental Biopsychiatry Research Program, McLean Hospital, 115 Mill Street, Belmont, Massachusetts 02478, USA \\
E-mail: lowen@mclean.org

\noindent
Bradley~M.~Jost \\
Air Force Research Laboratory, DELA, \\
3550 Aberdeen Avenue SE,
Kirtland AFB, New Mexico 87117, USA \\
E-mail: bjost@ieee.org

\noindent
Karin~Vibe-Rheymer \\
Department of Electrical and Computer Engineering, Boston University, \\
8 Saint Mary's Street, Boston, Massachusetts 02215-2421, USA \\
E-mail: kvibe@ieee.org

\noindent
Conor~Heneghan, \\
Department of Electronic and Electrical
Engineering, \\
University College Dublin, Belfield, Dublin 4, Ireland \\
E-mail: Conor.Heneghan@ucd.ie

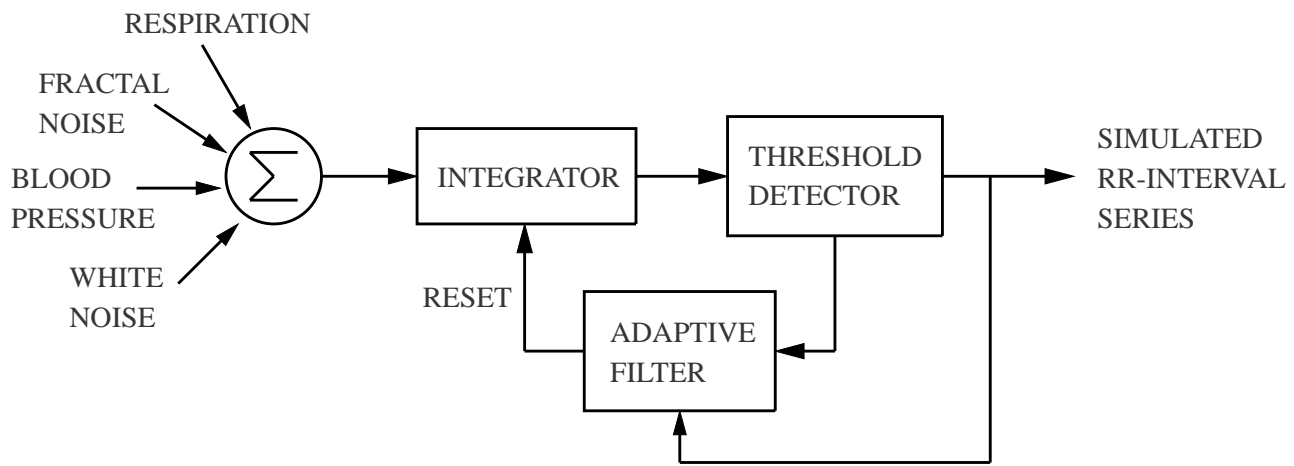
\end{document}